\documentclass[twoside,english,sort&compress]{iopart}
\usepackage[LGR,T1]{fontenc}
\usepackage[latin9]{inputenc}
\usepackage{geometry}
\usepackage{textcomp}
\usepackage{amstext}
\usepackage{graphicx}
\usepackage{esint}
\usepackage[numbers]{natbib}

\makeatletter

\DeclareRobustCommand{\greektext}{%
  \fontencoding{LGR}\selectfont\def\encodingdefault{LGR}}
\DeclareRobustCommand{\textgreek}[1]{\leavevmode{\greektext #1}}
\ProvideTextCommand{\~}{LGR}[1]{\char126#1}

\newcommand{\lyxmathsym}[1]{\ifmmode\begingroup\def\b@ld{bold}
  \text{\ifx\math@version\b@ld\bfseries\fi#1}\endgroup\else#1\fi}

\providecommand{\tabularnewline}{\\}

\usepackage{iopams}
\usepackage{setstack}

\newcommand{\eqref}[1]{(\ref{#1})}

\makeatother

\usepackage{babel}
\begin{document}
\title[Thermo-electromechanical coupling of BNT-type piezoelectric materials]{The influence of thermo-electromechanical coupling on the performance
of lead-free BNT-type piezoelectric materials}
\author{{\Large{}Akshayveer$^{1}$, Federico C Buroni$^{2}$, Roderick Melnik$^{1}$,
Luis Rodriguez-Tembleque$^{3}$, Andres Saez$^{3}$ and Sundeep Singh}$^{4}$}
\address{{\large{}$^{1}$MS2Discovery Interdisciplinary Research Institute,
Wilfrid Laurier University, Waterloo, Ontario N2L 3C5, Canada }}
\address{{\large{}$^{2}$Department of Mechanical Engineering and Manufacturing,
Universidad de Sevilla, Camino de los Descubrimientos s/n, Seville
E-41092, Spain}}
\address{{\large{}$^{3}$Department of Continuum Mechanics and Structural Analysis,
Universidad de Sevilla, Camino de los Descubrimientos s/n, Seville
E-41092, Spain}}
\address{{\large{}$^{4}$Faculty of Sustainable Design Engineering, University
of Prince Edward Island, Charlottetown, PEIC1A4P3, Canada}}
\ead{{\large{}aakshayveer@wlu.ca}}
\begin{abstract}
In recent times, there have been notable advancements in haptic technology,
particularly in screens found on mobile phones, laptops, LED screens,
and control panels. However, it is essential to note that the progress
in high-temperature haptic applications is still in the developmental
phase. Due to its complex phase and domain structures, lead-free piezoelectric
materials such as BNT-based haptic technology behave differently at
high temperatures than ambient conditions. Therefore, it is essential
to investigate the aspects of thermal management and thermal stability,
as temperature plays a vital role in the phase and domain transition
of BNT material. A two-dimensional thermo-electromechanical model
has been proposed in this study to analyze the thermal stability of
BNT material by analyzing the impact of temperature on effective electromechanical
properties and mechanical and electric field parameters. However,
the thermo-electromechanical modelling of BNT ceramics examines the
macroscopic effects of the applied thermal field on mechanical and
electric field parameters as phase change and microdomain dynamics
are not considered in this model. This study analyzes the impact of
thermo-electromechanical coupling on the performance of BNT-type piezoelectric
materials compared to conventional electromechanical coupling. The
results predicted a significant improvement in piezoelectric response
compared to electromechanical coupling due to increased thermoelectric
effect in absence of phase change and microdomain switching for temperature
boundary conditions below depolarization temperature ($T_{d}\sim200\lyxmathsym{\textcelsius}$
for pure BNT material).
\end{abstract}
\maketitle

\section{Introduction}

Haptic technology has become popular recently in various sensor and
actuator applications such as mobile phone screens, laptops, LED screens,
and different control panels \citep{Chen2023,Nguyen2023}. These devices
work on the sensors that sense the user's touch, the algorithms that
process the data, and the actuators that produce the haptic feedback
are all part of the machine side. The neuronal connections that provide
information to the brain and the haptic sensors found in the skin
make up the human side \citep{Krogmeier2019}. Various piezoelectric
materials are used in these haptic devices. Piezoelectric materials
transform mechanical energy into electrical energy and vice versa
\citep{Krishnaswamy2020d,Krishnaswamy2020a}. Furthermore, lead Zirconate
Titanate (PZT) is used in most piezoelectric applications due to its
ease of commercial access and potent piezoelectric response \citep{Ibn-Mohammed2017}.
PZT's central constituent element is lead, a poisonous chemical detrimental
to human health and more prone to environmental contamination \citep{Panda2009}.
Lead poisoning can lead to renal dysfunction, neural dysfunction,
defective reproduction system, weakening of bones and cellular dysfunction
in the human body \citep{Collin2022}. Furthermore, the level of lead
toxicity from PZTs increases with temperature \citep{Wu2003}. Therefore,
we must consider lead-free piezoelectric material options for haptic
technology, especially in high-temperature applications.

Lead-free piezoelectric composites offer scalable and eco-friendly
options for electromechanical actuators and mechanical energy harvesting
and sensing \citep{Ibn-Mohammed2017,Maurya2018,Tiller2019}. These
composites typically consist of relatively softer matrices, typically
polymers, filled with nano- or micro-particles of hard, crystalline,
lead-free piezoelectric materials with vigorous piezoelectric activity,
such as non-perovskites like bismuth layered structured ferroelectrics
(BLSF) and tungsten bronze ferroelectrics \citep{Takenaka2005}, and
perovskites like $BaTiO_{3}$ (BT), $Bi_{0.5}Na_{0.5}TiO_{3}$ (BNT),
and $KNaNbO_{3}$ (KNN) \citep{Shin2014,Quan2021,Zhang2023}. These
lead-free piezoelectric materials have lower piezoelectric coefficients,
but their piezoelectric performance may be enhanced by intrinsic (lattice
distortions) and extrinsic contributions (altering the domain topologies)
\citep{DraganDamjanovic1998}. By lattice softening and decreasing
unit cell distortion, a novel KNN-based piezoelectric material exhibits
a more robust piezoelectric response ($d_{33}\sim500pc/N$) and an
increased Curie temperature ($T_{c}\sim200\lyxmathsym{\textcelsius}$)
\citep{Liu2019}. In $(1-x)BaTiO_{3}-xCrZnO_{3}$, a remarkably high
piezoelectric response of ($d_{33}\sim445\pm20pc/N$) is produced
by controlling the multiphase coexistence of rhombohedral-orthorhombic-tetragonal
(R-O-T) and the phase transition temperature \citep{Wang2021}. Similarly,
0.94BNT-0.06BT exhibits a 200\% improvement in piezoelectric response
and enhanced thermal stability (depolarization temperature increases
to 57${^\circ}C$ from 32${^\circ}C$) by engineering template grain
formation utilizing NN templates, which harnessed the microstructure
of BNT-based ceramics \citep{Bai2018}.

However, BNT has a significantly more complicated domain structure
than BT and KNN due to its more complex phase structure. Pure BNT
has depicted moderate piezoelectric properties ($d_{33}\sim200pc/N$)
and high coercive electric field ($E_{c}\sim70kV/cm$) \citep{Hao2019}.
Pure BNT exhibits rhombohedral R3c, monoclinic Cc, and a mixture of
both phases below depolarization temperature ($T_{d}\sim200\lyxmathsym{\textcelsius}$)
depending on different electrical and mechanical treatments \citep{Rao2013,Gorfman2010}.
However, these existing phases exhibit ferroelectric (FE) domain switching
and show good piezoelectric properties. Above $T_{d}$, BNT exhibits
an orthorhombic Pnma phase with an anti-ferroelectric (AFE) character
between 200-320$\lyxmathsym{\textcelsius}$ \citep{Dorcet2008a,Dorcet2008,Dorcet2009,Trolliard2008}.
It is supposed to be a nonpolar or weakly-polar phase and exhibits
a maximum value of dielectric constant at 320$\lyxmathsym{\textcelsius}$
which is called the maximum dielectric temperature ($T_{m}$). Above
this temperature, BNT exhibits a paraelectric tetragonal P4bm phase
till 520$\lyxmathsym{\textcelsius}$ and convert to a symmetric cubic
Pm3m structure \citep{Geday2000}, this temperature is designated
as Curie temperature ($T_{c}$) above which piezoelectric effect is
negligible due to the symmetric nature of the cubic phase. As a result,
regulating these temperatures using various approaches can improve
the piezoelectric capabilities of BNT material. Various techniques,
including quenching, ceramic composites, binary or ternary solid solutions,
and ion replacement, can improve BNT's domain structure. These strategies
primarily influence domain structure, which in turn affects macro
performance of BNT-type piezoelectric materials. An adequate selection
of oxides for BNT/oxide composites may improve its thermal stability.
In $BNKT-Al_{2}O_{3}$, the depolarization temperature has been deferred
to higher temperatures (116\textcelsius -227$\lyxmathsym{\textcelsius}$).
The $BNKT-0.15Al_{2}O_{3}$ remains stable even at 210$\lyxmathsym{\textcelsius}$
\citep{Yin2018}. It was also found that the piezoelectric response
of 0.94BNT-0.06BT for thick films (285nm) increased with temperature
from 1100 to 1160$\lyxmathsym{\textcelsius}$. But it decreases at
1180$\lyxmathsym{\textcelsius}$ \citep{Liu2015a}. Considerable strain
and ultra-low hysteresis with suitable temperature and frequency stability
are achieved simultaneously in BNT-6BT-0.2SLT ceramic to highly dynamic
PNRs via pushing relaxor temperature to room temperature by SLT addition
to the BNT system \citep{Hou2022}.

Furthermore, with the inclusion of KNNG composition, $(1-x)BNT-xKNN$
$(x=0-0.02)$ and one wt\% Gd2O3 demonstrates increased piezoelectric
and dielectric capabilities by moving depolarization temperature ($T_{d}$)
and Curie temperature ($T_{c}$) values towards lower temperatures.
According to the findings, $BNT-xKNNG$ $(x=0.01)$ ceramics are promising
for high-power electromechanical applications \citep{Pattipaka2018}.
Furthermore, Jaroopam and Jaita \citep{Jarupoom2022} reported that
adding BMT to BNT increased grain size, dielectric, piezoelectric,
ferroelectric, and polarization characteristics owing to increased
maximum temperature ($T_{m}$). The ferroelectric ($P_{r}=23.84C/cm^{2},E_{c}=34.41kV/cm$)
and piezoelectric (low-field $d_{33}=159pC/N$) characteristics of
the$x=0.05$ in $(1-x)BNT-xBMT$ $(x=0-0.2)$ ceramic were both good.
Furthermore, including BMT increased the magnetic properties, energy
storage efficacy, and strain generated by electric fields for the
$x=0.05$ ceramic, which was approximately 3.4 times (240\%) more
effective than the pure BNT ceramic \citep{Jarupoom2022}. Mahdi and
Majid \citep{Mahdi2016} investigated the effect of polarization in
BNT-BKT-BT inclusion on the PVDF matrix in a hot press as a function
of volume percent. By increasing the volume percentage from 0 to 0.3,
the piezoelectric and pyroelectric coefficients rose from 28 to 40
pC/N and 26 to 95 $\mu Cm^{-2}K^{-1}$ correspondingly. A stable operating
temperature of 150$\lyxmathsym{\textcelsius}$ has been achieved by
changing the phase structure from R to RT coexisted phase. This is
achieved by adding PMN and BST to a non-stoichiometric BNT matrix,
making the material more ferroelectric relaxor by lowering the remnant
polarization and keeping the maximum polarization \citep{Yang2023}.
However, Zr-modified BNKT ceramics have been synthesized, and their
electromechanical properties are investigated. These results indicate
that an appropriate amount of Zr substitution significantly enhances
the field-induced strain level of BNKT ceramics, and higher values
of piezoelectric constant also increase \citep{Hussain2010}.

Moreover, in quenched BNT-7BT ceramics, this study has been shown
to raise the depolarization temperature ($T_{d}$) by 44 ${^\circ}C$
and achieve a temperature-independent $d_{33}$ throughout a broad
temperature range of 25\textendash 170 ${^\circ}C$ by improving the
ferroelectric state connected to the R3c phase \citep{Chen2022}.
However, adding 1 mol\% AlN in BNT raises $d_{33}$ from 165 to 234
pC/N and increases $T_{d}$ by 50 ${^\circ}C$ due to a rise in the
tetragonal phase. Furthermore, with this composition, the modified
ceramics have bigger grains and high-density lamellar nano-domains
with diameters ranging from 30 to 50 nm. Thus, polarization reversal
and domain mobility are greatly amplified, leading to the enormous
$d_{33}$. Temperature-dependent dielectric and XRD studies indicated
that the delayed thermal depolarization in the modified ceramics is
due to enhanced and poling-field stabilized tetragonal structure \citep{Zhou2022a}.
Furthermore, BF-BT-xBNT $(0<x<0.04)$ at $x=0.01$ exhibits the best
piezoelectric constant $d_{33}=206pC/N$ with $T_{c}=488{^\circ}C$.
It is produced in the sample due to the synergy effect of optimal
R/T phase ratio, enhanced tetragonality, increased density, and decreased
leakage current \citep{Cheng2023}.

Based on the aforementioned research, it can be inferred that optimizing
temperature management and thermal stability is crucial in improving
the piezoelectric properties of BNT and its substitute ceramics. Hence,
it is imperative to conduct a comprehensive investigation on the piezoelectric
properties of BNT under varying thermal circumstances since the material's
behaviour manifests distinct phases and domain structures at different
temperature levels. Most existing research on the investigating the
thermo-electromechanical behaviour of BNT materials is primarily experimental.
However, there is a need for more literature that specifically addresses
the numerical coupling of thermo-electromechanical phenomena. Numerical
studies have significant importance due to their cheap processing
cost and ability to accurately anticipate behaviour that closely aligns
with experimental observations. Experiments that are challenging to
conduct can be readily substituted by numerical simulations. There
is a limited amount of research available on the topic of thermo-electromechanical
coupling. While some studies have been undertaken on electromechanical
coupling \citep{Yang2006,Chaterjee2018,Krishnaswamy2020c}, it remains
challenging to comprehend the thermal stability and temperature regulation
using this type of modelling. Consequently, it is crucial to investigate
the thermo-electromechanical coupling of BNT material in order to
anticipate its behaviour under elevated temperatures. The BNT material
has intricate phase and domain structures, demonstrating heightened
complexity with increasing temperature. The investigation of phase
changes is vital for the advancement of high-temperature haptic applications.
Nevertheless, considering the intricate and extensive body of research
pertaining to the domain structure of ceramics based on BNT (bismuth
sodium titanate), our focus will be directed towards the alterations
occurring in the nano-domain as a result of external stimuli such
as composition, electric field, and temperature. Additionally, we
will examine the impact of these changes on the strain properties
at the macro-level and the fatigue behaviour of BNT-based ceramics.
Therefore, the primary objective of this study is to investigate the
thermo-electromechanical modelling of BNT ceramics, explicitly examining
the macroscopic impacts of phase changes on mechanical and electric
field parameters. However, the detailed modelling of domain switching
and the coexistence of phases are areas for further exploration in
this study.

\section{Mathematical model}

The steady-state behaviour of a two-dimensional piezoelectric composite
architecture comprised of microscale piezoelectric inclusions (BNT)
is investigated. The sub-sections that follow will go over the composite
architecture, the coupled thermo-piezoelectric model that was used
to study the composite's behaviour, the materials models that govern
the dielectric and mechanical properties of the composite, and the
boundary conditions that were used to compute specific effective electro-elastic
coefficients of interest.

\subsection{Geometrical description of piezoelectric composite and inclusions}

\begin{figure}
\includegraphics[scale=0.52]{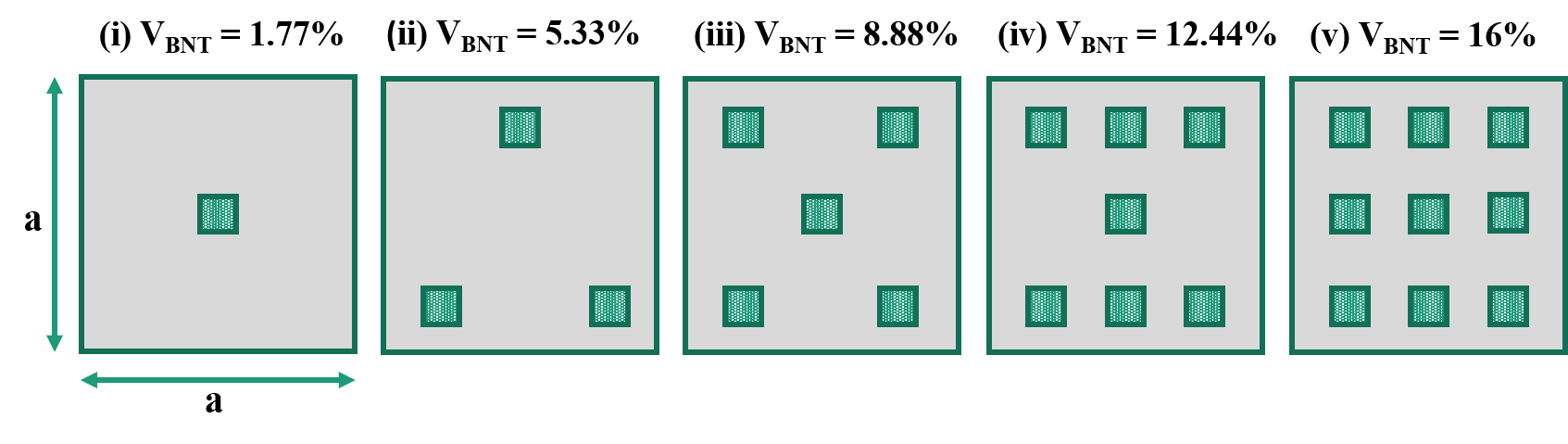}

\caption{Piezocomposites with progressive BNT inclusions with PDMS matrix.\label{fig:Schematic}}

\end{figure}

As illustrated schematically in Fig.\ref{fig:Schematic}, we represent
the composite as a two-dimensional RVE in the $x1-x3$ plane. The
composite comprises two parts: (i) the rectangular matrix with $a_{m}$
and $b_{m}$ dimensions, and (ii) microscale polycrystalline lead-free
BNT inclusions square shaped were included. The piezoelectric inclusions
are around 20 \textgreek{m}m in size, while the sides ($a_{m}$ and
$b_{m}$) of exemplified RVE geometry are 150 \textgreek{m}m long.
The geometric description is inspired by the work of Krishnaswamy
et al. \citep{Krishnaswamy2021}, who investigated same-sized matrix
geometry with a similar amount of BaTiO3 material inclusions. However,
they used random size inclusions, but we used squared shape BNT inclusions;
thus, the volume fraction is somewhat different for the same number
of inclusions. The volume fraction of BNT inclusions is dependent
on the number of inclusions, which is better described in Fig.\ref{fig:Schematic}.
The length scale of piezoelectric inclusions is microscopic, while
the length scale of the grain size is of nano-range. Therefore, the
consideration that must be taken into account is that the size of
the inclusions must be greater than the optimal grain size of the
composite to get enhanced piezoelectricity in the poly-crystal of
BNT.

\subsection{Coupled thermo-electromechanical model for piezoelectric composites}

The preceding section describes the composite geometry being examined
in this work. We describe the thermo-electromechanical model that
was used to investigate the composite design described in section
2.1 and schematically depicted in Fig.\ref{fig:Schematic}. In the
context of the linear theory, the Helmholtz free energy function $G$
considering three independent variables $(\varepsilon,E,\theta)$
of our system has been of the following form \citep{Chandrasekharaiah1988}:

\begin{eqnarray}
G(\varepsilon_{ij},E_{i},\theta) & = & 1/2C_{ijkl}\varepsilon_{ij}\varepsilon_{kl}-e_{ijk}E_{i}\varepsilon_{jk}-1/2\epsilon_{ij}E_{i}E_{j}\nonumber \\
 &  & -\mu_{ijkl}E_{i}\varepsilon_{jk.l}-\lambda_{ij}\theta\varepsilon_{ij}-\eta_{i}E_{i}\theta-\frac{1}{2}a_{T}\theta^{2},\label{eq:Eq1}
\end{eqnarray}

where$C_{ijkl}$ are the elastic constants, $\epsilon_{ij}$ are the
dielectric constants, $e_{ijk}$ are the piezoelectric constants,
$\mu_{ijkl}$ are the flexoelectric constants, $\lambda_{ij}$ are
the thermal modulus, $\eta_{i}$ is the thermoelectric constants,
and $a_{T}$ are heat capacity coefficients respectively. The heat
capacity coefficients $a_{T}$ are defined as $a_{T}=\frac{(C_{\varepsilon}^{V})}{T_{0}}$
where $C_{\varepsilon}^{V}$ is the heat capacity at constant volume
and elastic strain at ambient temperature $T_{0}$. The constitutive
relationships for thermo-electro-mechanical coupling are derived from
Eq.\ref{eq:Eq1} and are as follows \citep{Patil2009a}:

\begin{eqnarray}
\sigma_{ij}= & \frac{\partial G}{\partial\varepsilon_{ij}} & =c_{ijkl}\varepsilon_{kl}-e_{kij}E_{k}-\lambda_{ij}\theta,\label{eq:Eq2}
\end{eqnarray}

\begin{eqnarray}
\hat{\sigma_{ij}}= & \frac{\partial G}{\partial\varepsilon_{ij,k}} & =\mu_{ijkl}E_{l},\label{eq:Eq3}
\end{eqnarray}

\begin{eqnarray}
D_{i}= & \frac{\partial G}{\partial E_{i}} & =\epsilon_{ik}E_{k}+e_{ikl}\varepsilon_{kl}+\eta_{i}\theta-\mu_{klij}\varepsilon_{ij},\label{eq:Eq4}
\end{eqnarray}

\begin{eqnarray}
S= & \frac{\partial G}{\partial\theta} & =\lambda_{ij}\varepsilon_{ij}+\eta_{i}E_{i}+a_{T}\theta,\label{eq:Eq5}
\end{eqnarray}

where $S$ is the entropy and $\theta$ is the temperature difference.
The $\sigma_{ij}$ and $\varepsilon_{ij}$ are the elastic stress
and strain tensor components, respectively, whereas $D_{i}$ are the
electric flux density vector components, and $E_{k}$ are the electric
field vector components. Further, the governing equations that provide
equilibrium to all applied fields are as follows \citep{Patil2009a,Krishnaswamy2019}:

\begin{eqnarray}
(\sigma_{ij}-\hat{\sigma}_{ijk.k})_{j}+F_{i} & = & 0,\label{eq:Eq6}
\end{eqnarray}

\begin{eqnarray}
D_{i,i} & = & 0,\label{eq:Eq7}
\end{eqnarray}

\begin{eqnarray}
Q_{i,i}+Q_{gen} & = & 0,\label{eq:Eq8}
\end{eqnarray}

where $F_{i}$ denotes the components of body force expected to diminish.
In the absence of higher-order stress components $\hat{\sigma}_{ijk.k}$,
Eq.\ref{eq:Eq6} would reduce to linear momentum balance in the traditional
elastic formalism. The higher-order stresses may be read as moment
stresses, and the balance equation can be conceived as incorporating
the balance for linear and angular momentum. Eq.\ref{eq:Eq8} becomes
a simple steady-state heat conduction equation without heat generation
term $Q_{gen}$.

The gradient equations correspond to the relationships between the
linear strain and mechanical displacement, the electric field and
electric potential, and the thermal field and temperature change.
They are stated respectively as \citep{Krishnaswamy2020a,Patil2009a}:

\begin{eqnarray}
\varepsilon_{kl} & = & \frac{1}{2}(u_{k,l}+u_{l,k}),\label{eq:Eq9}
\end{eqnarray}

\begin{eqnarray}
E_{k} & = & -V,_{k},\label{eq:Eq10}
\end{eqnarray}

\begin{equation}
Q_{i}=-k\theta,_{i},\label{eq:Eq11}
\end{equation}

where $\varepsilon_{kl}$, $E_{k}$, $Q_{i}$, $u$, $V$, $k$, and
$\theta$ are the components of the strain tensor, electric field
vector, thermal field vector, mechanical displacement vector, electric
potential, thermal conductivity and temperature change from the reference,
respectively.

\subsection{Piezoelectric response of composites and boundary conditions}

\begin{figure}
\includegraphics[scale=0.55]{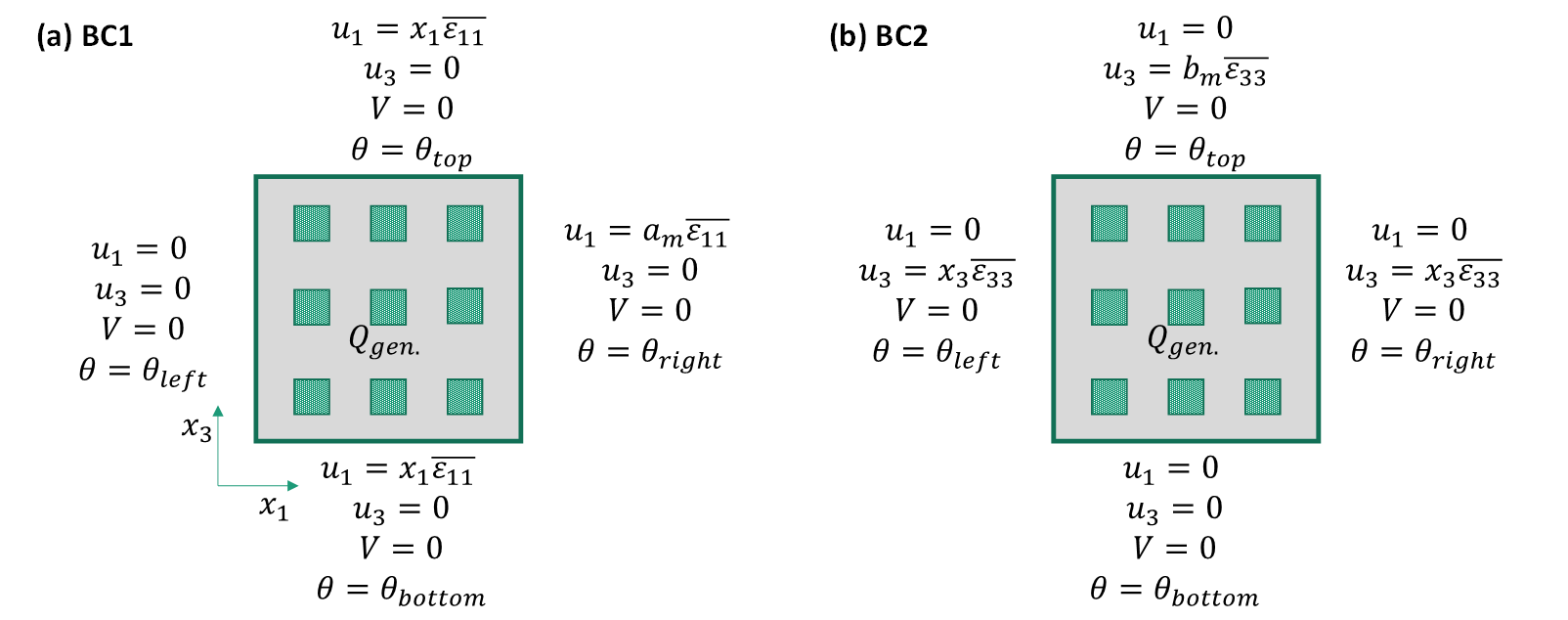}

\caption{Different boundary conditions for thermo-electromechanical coupling
of BNT-type piezoelectric material with PDMS matrix \citep{Krishnaswamy2019}.\label{fig:Different-boundary-conditions}}

\end{figure}

The previous sections described the features of the composite geometry
explored here and the coupled equations that regulate the composite's
piezoelectric and thermoelectric behaviour. We compute the effective
$e_{31}$ and $e_{33}$ of the composite to evaluate its piezoelectric
response. These effective coefficients must be determined using two
independent boundary conditions that apply axial strains along the
$x_{1}$ and $x_{3}$ directions BC1 and BC 2, as described in Fig.
\ref{fig:Different-boundary-conditions}(a) and \ref{fig:Different-boundary-conditions}(b),
respectively.

These two different mechanical boundary conditions are analyzed for
different thermal BCs, as other thermal conditions can have different
thermo-elastic and thermoelectric behaviour. The different types of
thermal boundary conditions for BC1 and BC2 are as follows:
\begin{itemize}
\item One side heating: 
\begin{itemize}
\item (a) BC1: $\theta_{right}=0-150\lyxmathsym{\textcelsius}$, $\theta_{left}=\theta_{top}=\theta_{bottom}=0$, 
\item (b) BC2: $\theta_{top}=0-150\lyxmathsym{\textcelsius}$, $\theta_{left}=\theta_{right}=\theta_{bottom}=0$,
and $Q_{gen.}=0$ for both BC1 and BC2. 
\end{itemize}
\item Two side heating: 
\begin{itemize}
\item (a) Adjacent wall heating: $\theta_{top}=\theta_{right}=0-150\lyxmathsym{\textcelsius}$,
$\theta_{left}=\theta_{bottom}=0$ and $Q_{gen.}=0$ for both BC1
and BC2. 
\item (b) Opposite wall heating: 
\begin{itemize}
\item (a) BC1:$\theta_{left}=\theta_{right}=0-150\lyxmathsym{\textcelsius}$,
$\theta_{top}=\theta_{bottom}=0$, 
\item (b) BC2: \textgreek{j}\_\{top\} = $\theta_{bottom}=0-150\lyxmathsym{\textcelsius}$,
$\theta_{left}=\theta_{right}=0$, and $Q_{gen.}=0$ for both BC1
and BC2. 
\end{itemize}
\end{itemize}
\item Three side heating: 
\begin{itemize}
\item (a) BC1: $\theta_{left}=0$, $\theta_{right}=\theta_{top}=\theta_{bottom}=0-150\lyxmathsym{\textcelsius}$, 
\item (b) BC2: $\theta_{bottom}=0$, $\theta_{left}=\theta_{right}=\theta_{top}=0-150\lyxmathsym{\textcelsius}$,
and $Q_{gen.}=0$ for both BC1 and BC2. 
\end{itemize}
\item All side heating: $\theta_{left}=\theta_{right}=\theta_{top}=\theta_{bottom}=0-150\lyxmathsym{\textcelsius}$,
and $Q_{gen.}=0$ for both BC1 and BC2. 
\item Internal heat generation: $\theta_{left}=\theta_{right}=\theta_{top}=\theta_{bottom}=0\lyxmathsym{\textcelsius}$,
and $Q_{gen.}=10^{7}-10^{11}$ $W/m^{3}$for both BC1 and BC2. 
\end{itemize}
Using boundary conditions BC1, we determine the composite's effective
characteristics $e_{11,eff.}$,$c_{11,eff.}$, and $c_{13,eff.}$
and, using the second set of boundary conditions BC2, we determine
the effective characteristics $e_{33,eff.}$, $c_{33,eff.}$, and
$c_{13,eff.}$ of the composites. The Voigt notation is used to convert
four- and three-digit indices of elastic coefficients, and piezoelectric
coefficients into two-index notation for simplifying the notation
system and reducing the number of variables to reduce the computational
cost. The volume average of quantity $A$ is denoted as $\left\langle A\right\rangle $
in the following calculation. It is computed as \citep{Krishnaswamy2020d,Saputra2017}:

\begin{eqnarray}
\left\langle A\right\rangle  & = & \frac{1}{(a_{m}b_{m})}\int_{\Omega}Ad\Omega,\label{eq:Eq12}
\end{eqnarray}

Where \textgreek{W} is the volume across which the integration is
performed, the complete volume of the RVE in this example. The following
volume averages are derived by applying the boundary criteria BC1
\citep{Krishnaswamy2020d,Saputra2017} where the applied axial strain
in the x-axis $(\varepsilon_{11})$ is equal to the average axial
strain $(\bar{\varepsilon}_{11})$ in the x-axis, while the applied
axial strain in the y-axis $(\varepsilon_{33})$ and the applied transverse
strain $(\varepsilon_{13})$ is zero. Also, the applied electric field
$(E_{33})$ in the poling axis (y-axis) is zero, as shown in Eq. \ref{eq:Eq13}:

\begin{eqnarray}
\varepsilon_{11} & = & \bar{\varepsilon}_{11},\varepsilon_{33}=0,\varepsilon_{13}=0,E_{33}=0\label{eq:Eq13}
\end{eqnarray}

Using these volume averages, the composite's effective coefficients
are derived as follows\citep{Krishnaswamy2020d,Saputra2017}:

\begin{eqnarray}
\text{\text{\ensuremath{e_{31,eff.}}=\ensuremath{\frac{\left\langle D_{3}\right\rangle }{\varepsilon_{11}}}},} & c_{11,eff.}=\frac{\left\langle \sigma_{11}\right\rangle }{\varepsilon_{11}}, & c_{13,eff.}=\frac{\left\langle \sigma_{33}\right\rangle }{\varepsilon_{11}},\label{eq:Eq14}
\end{eqnarray}

where $\left\langle D_{3}\right\rangle $ is the volume average of
the electric flux density vector's $D_{3}$ component. Similarly,
using the second set of boundary conditions BC2, the volume averages
shown below are produced \citep{Krishnaswamy2020d,Saputra2017} where
the applied axial strain in the y-axis $(\varepsilon_{33})$ is equal
to the average axial strain $(\varepsilon_{33})$ in the y-axis, while
the applied axial strain in the x-axis $(\varepsilon_{11})$ and the
applied transverse strain $(\varepsilon_{13})$ is zero. Also, the
applied electric field $(E_{33})$ in the poling axis (y-axis) is
zero, as shown in Eq.\ref{eq:Eq15}:

\begin{eqnarray}
\varepsilon_{11} & =0, & \varepsilon_{33}=\bar{\varepsilon}_{33},\varepsilon_{13}=0,E_{33}=0\label{eq:Eq15}
\end{eqnarray}

These volume averages are then utilized to compute the composite's
effective coefficients \citep{Krishnaswamy2020d,Saputra2017}:

\begin{eqnarray}
\text{\text{\ensuremath{e_{33,eff.}}=\ensuremath{\frac{\left\langle D_{3}\right\rangle }{\varepsilon_{33}}}},} & c_{33,eff.}=\frac{\left\langle \sigma_{33}\right\rangle }{\varepsilon_{33}}, & c_{13,eff.}=\frac{\left\langle \sigma_{11}\right\rangle }{\varepsilon_{33}},\label{eq:Eq16}
\end{eqnarray}

We assume the development of minor values of axial stresses inside
composites due to applied axial strains in our computations and set
$\varepsilon_{11}$ and $\varepsilon_{33}$ in BC1 and BC2, respectively,
to $1\times10^{-6}$. The finite element analysis, performed using
COMSOL Multiphysics software, analyzes the behaviour of the composites
based on the thermo-electromechanical model and the boundary conditions
placed on the geometrically tailored composite architecture.

\subsection{Material Properties}

For our investigation, we settled on polydimethylsiloxane (PDMS),
a soft polymer matrix. The literature has been paying more and more
attention to experimental efforts to create piezoelectric composites
with soft matrices through 3D printing and other cutting-edge technologies.
These soft matrices provide two challenges, however. Initially, the
applied mechanical stimuli from the BNT are screened by their smooth
elastic characteristics, which leads to a relatively minimal production
of electric flux. Second, the passage of electric change out of the
piezoelectric inclusions needs to be improved by the typical weakness
of polymeric materials such as dielectrics. Nonetheless, other tests,
including recent initiatives in these areas, suggest that getting
around both problems above is possible. Based on these experimental
findings, a soft matrix, represented by PDMS in this instance, is
a prime choice for assessing the potential for elastic and dielectric
enhancements to increase the piezoelectric response. The material
constants of the PDMS matrix and BNT inclusions for which computations
have been performed are compiled in Table \ref{tab:Material-Prop}.
Analysis of the thermal stability and degradation of polymeric composites
based on PDMS is a crucial problem which should be examined for higher
temperatures.

\begin{table}[hb]
\caption{\label{tab:Material-Prop}Material properties required for the simulations.}

\centering{}%
\begin{tabular}{|@{}l|l|l|}
\hline 
\textbf{ Material Property} & \textbf{Values of BNT} & \textbf{Values of PDMS Matrix \citep{Krishnaswamy2020}}\tabularnewline
\hline 
\textbf{Elastic coefficients (GPa):} &  & \tabularnewline
$c_{11}$  & 153.9 \citep{Bujakiewicz-Koronska2008} & $\lambda_{m}+2\mu_{m}$\tabularnewline
$c_{13}$ & 52.1 \citep{Bujakiewicz-Koronska2008} & $\lambda_{m}$\tabularnewline
$c_{33}$  & 168.1 \citep{Bujakiewicz-Koronska2008} & $\lambda_{m}+2\mu_{m}$\tabularnewline
$c_{44}$  & 82.3 \citep{Bujakiewicz-Koronska2008} & $\mu_{m}$\tabularnewline
Young\textquoteright s Modulus ($E_{m}$)  & 93.0 \citep{JenniferAnneHaley2001} & 0.002\tabularnewline
Poisson\textquoteright s Ratio ($\nu_{m}$) & 0.23 \citep{JenniferAnneHaley2001} & 0.499\tabularnewline
\hline 
\textbf{Relative permittivity: } &  & \tabularnewline
$\frac{\epsilon_{11}}{\epsilon_{0}}$ & 367 \citep{Hiruma2009} & 2.72\tabularnewline
$\frac{\epsilon_{33}}{\epsilon_{0}}$ & 343 \citep{Hiruma2009} & 2.72\tabularnewline
\hline 
\textbf{Piezoelectric coefficients $(pC/N)$:}  &  & \tabularnewline
$d_{15}$  & 87.3 \citep{Bujakiewicz-Koronska2008} & Non-piezoelectric\tabularnewline
$d_{31}$  & -15.0 \citep{Bujakiewicz-Koronska2008} & \tabularnewline
$d_{33}$ & 72.9 \citep{Bujakiewicz-Koronska2008} & \tabularnewline
\hline 
\textbf{Flexoelectric coefficients $(cm^{-1})$:} &  & \tabularnewline
$\mu_{11}$, longitudinal  & $10^{-6}$\citep{Bujakiewicz-Koronska2008} & $10^{-9}$\tabularnewline
$\mu_{12}$, transverse  & $10^{-6}$\citep{Bujakiewicz-Koronska2008} & $10^{-9}$\tabularnewline
$\mu_{44}$, shear  & 0 \citep{Bujakiewicz-Koronska2008} & 0\tabularnewline
\hline 
\textbf{Thermo-elastic coefficients:} &  & \tabularnewline
$\lambda_{ij}$ & $2\times10^{-5}$.$K^{-1}$\citep{Suchanicz2010} & 3.$2\times10^{-4}$$K^{-1}$\tabularnewline
\hline 
\textbf{Thermoelectric coefficient: } &  & \tabularnewline
$\eta_{i}$  & $27.3\times10^{-4}$$C/m^{2}K$ \citep{Liu2015a}. & Non-thermoelectric\tabularnewline
\hline 
\textbf{Heat capacity coefficient:}  &  & \tabularnewline
$a_{T}$ & 500 $J/kgK$ \citep{Liu2015a}. & 1460 $J/kgK.$\tabularnewline
\hline 
\end{tabular} 
\end{table}

\section{Results and discussions}

\subsection{Effects of thermal boundary conditions on mechanical field}

The mechanical domain is influenced by the implementation of thermal
boundary conditions, resulting in the emergence of thermal stress
and strain. These factors subsequently induce modifications in a range
of mechanical and elastic characteristics. The creation of heat stress
is impacting the effective directional elastic coefficients. In addition,
heat stress and strain impact both the primary and transverse mechanical
strain. Furthermore, it should be noted that the arrangement of piezoelectric
inclusions has a significant effect on specific mechanical characteristics,
including primary and transverse strain. The amplitude and distribution
of thermal stress and strain are influenced by several types of thermal
boundary conditions, including one-side heating, two-side heating,
three-side heating, all-side heating, and internal heat generation.

\subsubsection{Effects of thermal boundary conditions on mechanical field}

\begin{figure}
\includegraphics[scale=0.5]{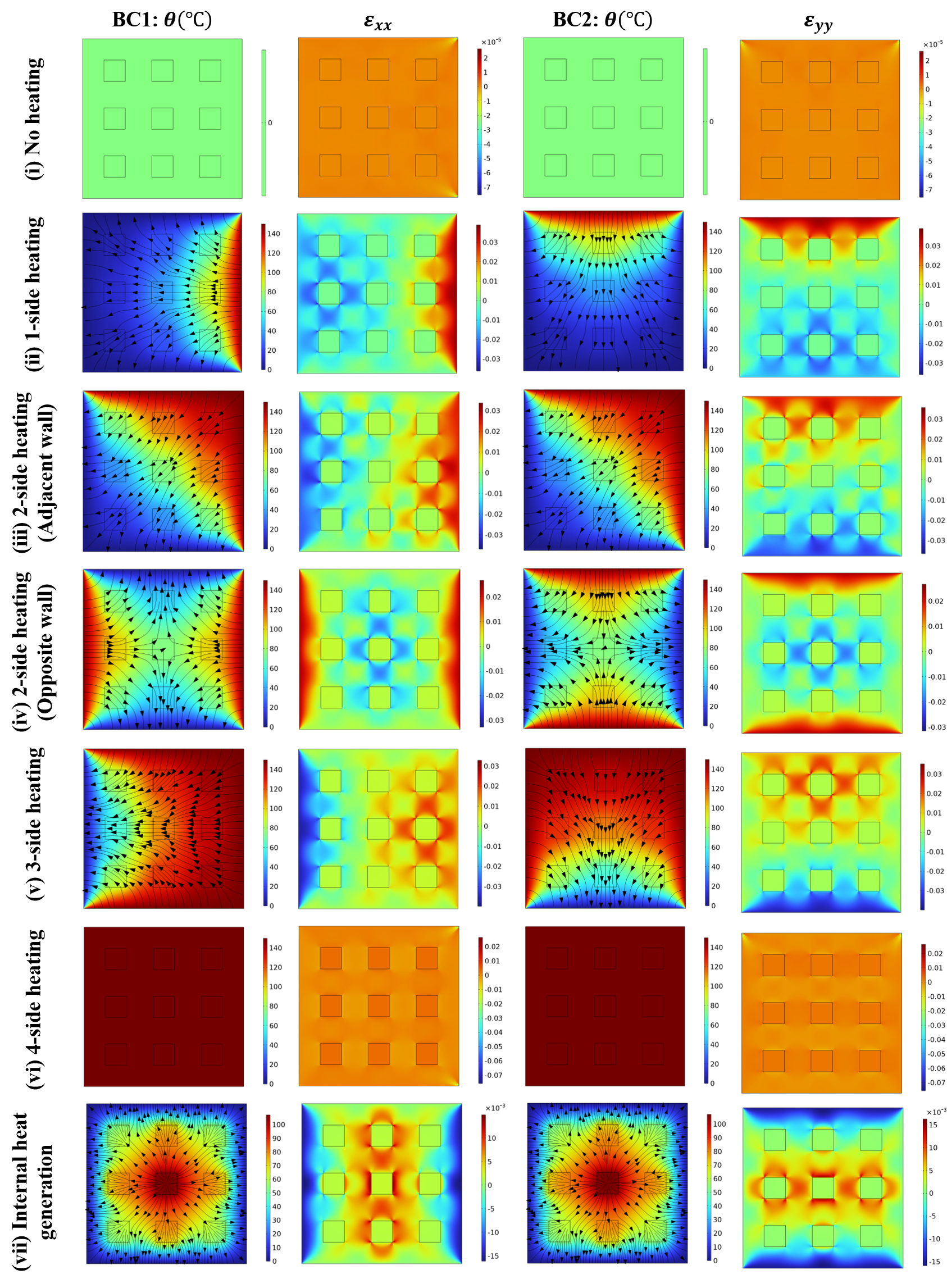}

\caption{Temperature and strain distribution contours for different types of
external heating (applied $\theta=150\lyxmathsym{\textcelsius}$)
and internal heat generation ($Q_{gen.}=10^{11}W/m^{3}$) for $V_{BNT}=16$\%
at mechanical boundary conditions BC1 \& BC2. Note that the vector
lines show the direction of heat transfer.\label{fig:Temperature-and-strain}}

\end{figure}

The piezoelectric-PDMS matrix will respond differently depending on
the temperature and mechanical boundary conditions. The thermal field
will alter the mechanical field due to the development of thermal
stresses and strains, which will also affect the piezoelectric performance
of the materials. Changing the amplitude and direction of the thermal
boundaries will influence the heat transfer rate and temperature distribution.
Furthermore, changes in the number of thermal constraints will affect
the mechanical field. As a result, research into the mechanical behaviour
of BNT-based piezoelectric composites at various thermal boundary
conditions is urgently needed. Fig. \ref{fig:Temperature-and-strain}
depicts the temperature and strain distribution contours in the different
boundary conditions. The heat transfer lines are positioned on the
temperature distribution curve, indicating the direction of heat transfer
and heat flow in the matrix and the inclusions. When these lines travel
into inclusions, their slope gets steeper. As a result, the heat transmission
rate in inclusions is more significant than in the matrix, and the
temperature gradient is reduced, resulting in a relatively uniform
temperature within the inclusions. However, the temperature gradient
in the matrix and piezo-inclusions in the direction of heat transport
may be visualized, which will govern the creation of thermal stresses
and strains in the composite. Since the distribution of inclusions
inside the composite varies with volume fraction, and both inclusions
and matrix display distinct mechanical properties, the stress and
strain distributions will be influenced by inclusion orientation in
the composite.

As illustrated in Fig. \ref{fig:Temperature-and-strain}(i), the strain
distribution relies purely on mechanical stress until there is no
external or internal heating of the geometry. The strain is equally
distributed throughout the BNT-based piezoelectric composite and only
reaches high values at the boundary corner when mechanical force is
applied. The use of thermal boundary conditions increased the amount
of the strain owing to thermal strain induction, which will help mechanical
strain in the direction of the applied load. In Fig. \ref{fig:Temperature-and-strain}(ii),
the geometry is heated on one side, and a thermal boundary condition
is imposed on the exact boundary where mechanical loading is applied.
At the same time, all other boundaries remain at ambient temperature.
The thermal boundary can expand in response to applied mechanical
and thermal loading, whereas the opposite edge to the applied thermal
boundary is restricted to any displacement or size change. As a result,
it is evident that the thermal strain is most significant close to
the thermal border and diminishes in all other directions in proportion
to the decrease in temperature gradient. Because the piezoelectric
inclusions have smaller dimensions, lower temperature gradients, and
more robust elastic characteristics than the matrix, their strain
distribution is more uniform. Since the wall across from it is immovable,
compressive thermal stress is generated close to it, causing negative
strain in this area. Using the thermal boundary condition boosted
strain production inside the matrix and the piezoelectric effect along
the inclusions, as explained further in section 3.2 of the study.
As a result, increasing the heat strain benefits piezoelectric behaviour
until the material breaks mechanically.

Increasing the thermal boundaries improve the strain even further.
In this situation, the wall with mechanical loading and its neighbouring
wall is heated, while the other two walls remain at ambient temperature
(see Fig. \ref{fig:Temperature-and-strain}(iii)). The amount and
distribution of thermal strain grow as the high-temperature zone spreads.
Maximum thermal strain is detected at the boundary where mechanical
loading is applied, whereas higher compressive strain is recorded
in the region of the mechanically constrained border. The propagation
of longitudinal stresses is reduced from mechanical loading borders
to mechanical constraint boundaries. The temperature distribution
is such that thermal stress affects the longitudinal and transverse
strains, resulting in the strain distribution pattern seen above.
The thermal strain at the neighbouring wall was expected to be larger;
however, the result is similar due to the production of compressive
stresses. 

Furthermore, the BNT-type piezoelectric composite is investigated
for two opposite wall heating sides, which display a distinct strain
distribution pattern, as illustrated in Fig. \ref{fig:Temperature-and-strain}
(iv). The mechanical loading and constraint borders are subjected
to thermal boundary conditions, while the other two walls remain at
room temperature. Inside the geometry, the strain distribution follows
the temperature distribution. The heat transfer direction is from
heated walls to cold walls, and it is evident that heat transmission
at the geometry's core is zero.

Similarly, the longitudinal strain distribution is excellent at mechanical
loading and limited boundaries. It decreases as we travel away from
the walls owing to thermal strain formation, and the centre of geometry
is dictated by compressive strain caused by thermal stress. However,
the strain distribution for all piezo-inclusions is approximately
uniform in this type of thermal loading, which can be a good factor
for a uniform piezoelectric response. However, with this thermal loading,
the strain distribution for all piezo-inclusions is nearly uniform,
which might be a beneficial factor for consistent piezoelectric response.

Increasing the number of thermal barriers to three further spreads
the thermal strain distribution, as illustrated in Fig. \ref{fig:Temperature-and-strain}
(v). In this circumstance, all three walls except the mechanical constraint
boundary are heated, and heat transfer is from these three walls to
the mechanical constraint border. As a result of the mechanical strain
generated by compressive thermal stress, the strain distribution is
highest at the mechanical loading boundary. It decreases in all other
directions, becoming negative in the proximity of the mechanically
constraint boundary. However, under these types of thermal settings,
the spread of thermal strain has been enhanced further and is determined
by temperature dispersion inside the geometry. Furthermore, the number
of thermal barriers has been extended to four in Fig. \ref{fig:Temperature-and-strain}
(vi), and it is discovered that no temperature gradient is noticed
inside the geometry. However, the temperature of the entire geometry
rises due to heating. Across the whole geometry, a uniform yet increased
strain distribution was observed. Only strain fluctuation at the corners
of the mechanical loading boundary is noticed, which is caused by
mechanical load. The strain elevation is caused by thermal pressure
created by all side heating, and thermal stress creation is low since
the temperature gradient is slight in the geometry. Uniform thermal
strain distribution also implies that displacement owing to temperature
rise is constant in all directions, implying that displacement variance
is attributable only to mechanical stress. The whole composite is
designed to have the same piezoelectric response across all inclusions,
although this might vary owing to mechanical loading.

We have previously investigated the influence of non-uniform and uniform
external heating on the BNT-based piezoelectric composite, and it
has been discovered that a rise in thermal barriers increases the
spread of thermal strain distribution inside the composite, which
becomes uniform in the case of constant heating. This is believed
to improve piezoelectric performance; however, the emphasis on internal
heat production in the mechanical field needs to be thoroughly examined,
as deduced from Fig. \ref{fig:Temperature-and-strain} (vii). All
boundaries are held at room temperature, and each piezoelectric inclusion
undergoes internal heat production. The heat transmission is uniformly
spreading outwards from the source to the walls, and the temperature
distribution is identical, more significant in the centre and decreasing
uniformly towards the wall; nevertheless, heat creation from each
inclusion raises the temperature at that specific place. The strain
distribution is also the same, with a maximum in the centre and a
decrease towards the wall. However, compressive mechanical strain
is detected near the mechanical constraint and loading border. The
most essential aspect is that all inclusions are subjected to increased
thermal strain, which is expected to improve piezoelectric performance.

It can be assumed that external or internal heating will favourably
impact the mechanical field and cause increased longitudinal strain,
which will aid in obtaining an improved piezoelectric response. However,
consistent heating on all sides provides a uniform, improved mechanical
field, which is advantageous in upgrading uniform piezoelectric response.

\subsubsection{Effects of thermal boundary conditions on effective elastic coefficients
of materials:}

\begin{figure}
\includegraphics[scale=0.15]{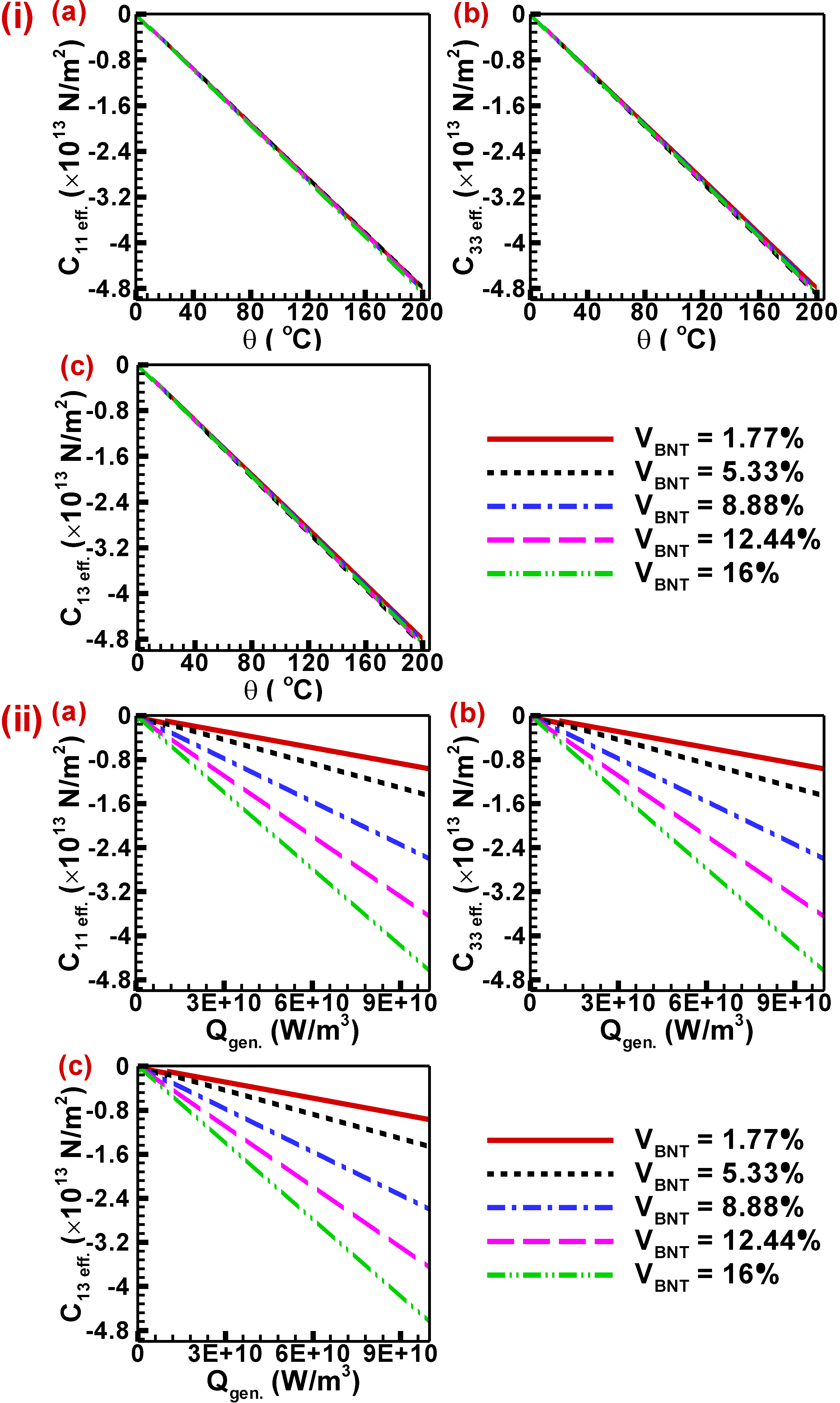}

\caption{(i) Effect of one side heating on effective elastic properties such
as (a) $C_{11eff.}$, (b) $C_{33eff.}$, and (c) $C_{13eff.}$. (ii)
Effect of internal heat generation on effective elastic properties
such as (a) $C_{11eff.}$, (b) $C_{33eff.}$, and (c) $C_{13eff.}$\label{fig:(i)-Elastic-coeff}}

\end{figure}

The calculation of effective elastic coefficients is based on Eqs.
\ref{eq:Eq14} and \ref{eq:Eq16}, which demonstrate that these coefficients
are exclusively determined by the major plane stresses. These stresses,
in turn, are influenced by the thermal stresses that are created.
The thermal compressive stresses are generated solely in principal
planes as a result of heating. Consequently, the impact of this phenomenon
on the effective elastic coefficients of the material is observed
in a similar manner. The effective elastic coefficients $C_{11eff.}$
and $C_{33eff.}$ in the principal directions are influenced by compressive
stress. However, the effective elastic coefficient $C_{13eff.}$ is
also affected by temperature boundary conditions as $C_{13eff.}$
is defined as the ratio of the average principal stress in one direction
to the average principal strain in the other direction. The thermal
compressive stress is the primary contributing factor to the variation
of the principal stress and effective elastic coefficients. Fig. \ref{fig:(i)-Elastic-coeff}
(i) illustrates the relationship between temperature and the fluctuation
of effective elastic coefficients in the case of one-side heating.
In the absence of heating, the effective elastic coefficients exhibit
a positive behaviour attributed to mechanical constraints, resulting
in a rise in their values with an increasing volume percentage of
piezoelectric inclusions. As the temperature rises, the dominance
of thermal compressive stresses over tensile mechanical stress increases,
leading to negative values for the overall effective elastic coefficients.
The influence of the volume fraction becomes negligible for the effective
elastic coefficients $C_{11eff.}$, $C_{13eff.}$, and $C_{33eff.}$
due to the compensatory effect of thermal compressive stress in both
principal planes, which counteracts the rise in tensile stress resulting
from mechanical constraint. However, the average strain in both primary
directions is essentially equal due to identical thermal and mechanical
boundary conditions. Furthermore, the elastic and thermo-elastic properties
of BNT in these principal directions demonstrate a close resemblance,
respectively. The BNT possesses similar elastic and thermo-elastic
characteristics in their primary directions, is subjected to similar
mechanical and thermal boundary conditions, and will experience comparable
mechanical and thermal stresses, respectively. As a result, the net
average strain in both directions displays minimal disparity. This
observation is supported by Fig. \ref{fig:(i)-Elastic-coeff} (i),
which indicates that all effective coefficients demonstrate comparable
trends as the wall temperature increases. Additionally, the influence
of increasing volume fraction and distribution of inclusions becomes
negligible under thermal loading conditions, as the change in effective
coefficients with volume fraction is minimal.

Furthermore, the relationship between effective elastic coefficients
and internal heat production is depicted in Fig. \ref{fig:(i)-Elastic-coeff}
(ii). At the outset, the effective elastic coefficients exhibit positive
values without any thermal boundary conditions. The presence of internal
heat generation leads to the development of thermal compressive stress
in the region of a piezoelectric inclusion. This stress has the effect
of reducing the effective elastic coefficients. The effective elastic
coefficients have greater negative values due to the dominance of
thermal compression stress over mechanical tensile stress. All elastic
coefficients display a similar pattern, as they are all determined
by the primary stresses and strains, as defined in Eqs. \ref{eq:Eq14}
and \ref{eq:Eq16}. The thermal compressive stress is dependent on
the volume percentage of piezoelectric inclusions as a result of internal
heat production inside the BNT-type piezoelectric material. In instances
where there is internal heat generation, the thermal compressive stress
experiences a rise in magnitude as the volume portion also increases.
Consequently, the effective elastic coefficients exhibit an upward
trend in negative values with the escalation of the volume \% of piezo-inclusions.
Moreover, the negative values of elastic coefficients become more
pronounced when internal heat generation increases due to the corresponding
rise in temperature and the resulting thermal compressive stress.
The relationship between effective elastic coefficients and heating
is consistent in both internal and external heating scenarios. However,
the relationship between effective elastic coefficients and volume
fraction differs between the two scenarios. This difference arises
from the fact that each piezo-inclusion is individually heated in
the internal heating scenario, but in the external heating scenario,
this is not the case.

\begin{figure}
\includegraphics[scale=0.15]{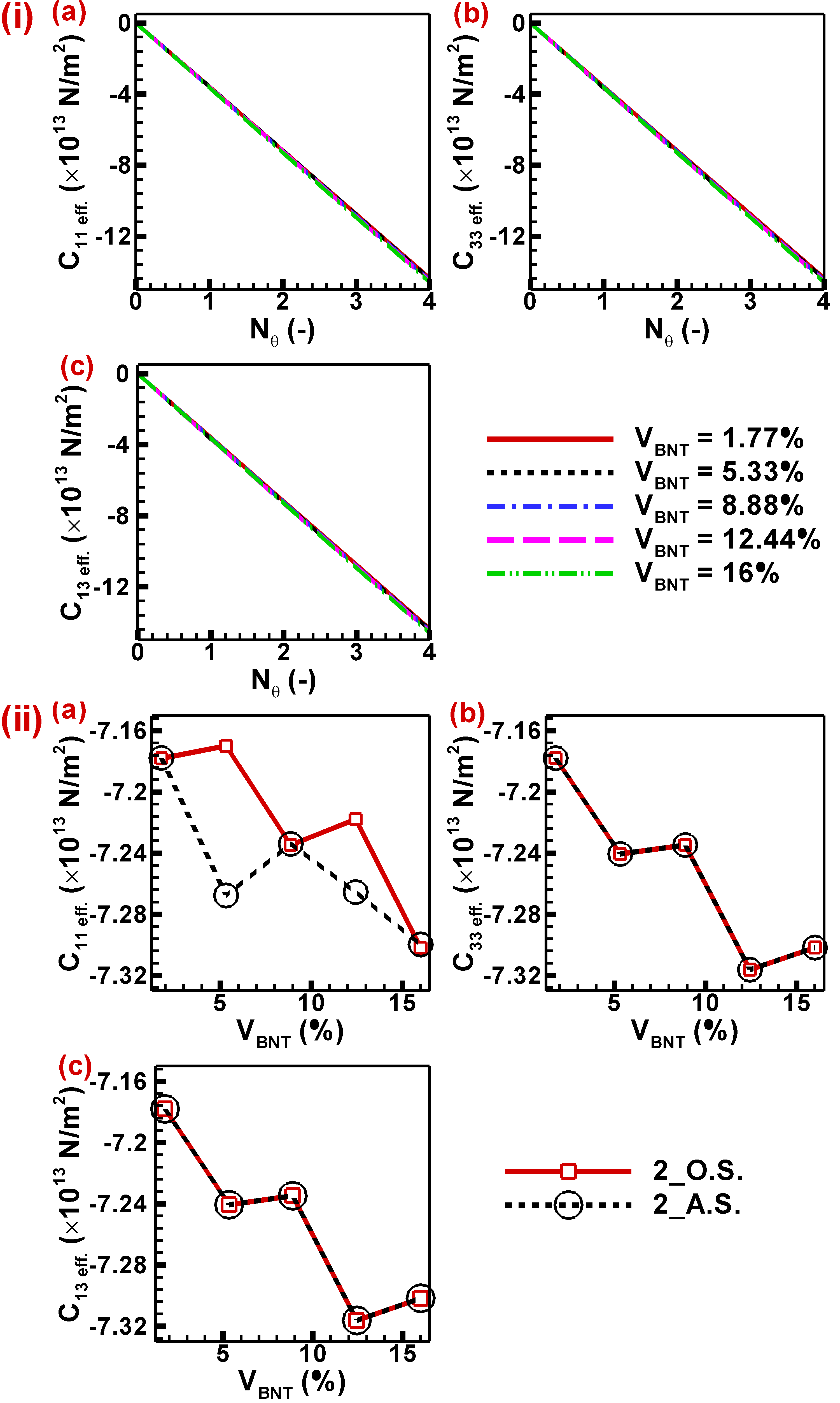}

\caption{(i) Effect of increase in the number of external thermal boundaries
on effective elastic properties such as (a) $C_{11eff.}$, (b) $C_{33eff.}$,
and (c) $C_{13eff.}$. (ii) Comparison of effective elastic properties
such s (a) $C_{11eff.}$, (b) $C_{33eff.}$, and (c) $C_{13eff.}$
for two sides adjacent wall heating and opposite wall heating.\label{fig:(i)-TB-elastic}}

\end{figure}

Further, Fig. \ref{fig:(i)-TB-elastic} (i) illustrates the relationship
between the effective elastic coefficients and the growing number
of thermal boundaries. The statement suggests that a growing number
of boundaries refers to the expansion of the boundaries where thermal
boundary conditions are being implemented. A value of 0 indicates
the absence of heating, signifying the presence of electromechanical
coupling exclusively. On the other hand, values of 1, 2, 3, and 4
correspond to one-side heating, two-side heating, three-side heating,
and all-side heating, respectively. It is evident that the effective
elastic coefficient demonstrates positive values in the absence of
heating, as it is subjected solely to tensile mechanical stress. Furthermore,
these values indicate an upward trend with the augmentation of the
volume fraction of BNT inclusions. However, the inclusion of thermal
boundary conditions on a single boundary, namely one-side heating,
results in the generation of thermal stress (compressive in nature)
that exceeds the mechanical tensile stress. As a consequence, the
effective elastic coefficients exhibit negative values. The significance
of increasing the volume fraction diminishes as it leads to an increase
in the magnitude of tensile stress created. Simultaneously, there
is a proportional rise in thermal compressive stress, resulting in
roughly equivalent and negative values of effective elastic coefficients.
Furthermore, the augmentation of thermal barriers results in a heightened
magnitude of thermal stress. Consequently, the net principal stress
becomes more compressive, leading to a rise in its negative value.
This finding suggests that the effective elastic coefficients have
a greater magnitude of negative values as the temperature bounds grow,
and this relationship follows a linear trend. The significance of
increasing the volume fraction diminishes as it leads to an increase
in the magnitude of produced tensile stress. Simultaneously, there
is a proportional rise in thermal compressive stress, resulting in
roughly equivalent and negative values of effective elastic coefficients.
Furthermore, the amplification of thermal stress is observed when
the number of thermal boundaries is increased. This results in a greater
compressive net principal stress, leading to a rise in its negative
value. This suggests that the effective elastic coefficients have
a greater magnitude of negative values as the temperature bounds grow,
and this relationship follows a linear trend.

Moreover, the direction of piezoelectric inclusions assumes a significant
function when the orientation of heating is altered in the context
of two-sided heating. In the context of two-sided heating, two instances
are often examined: neighboring wall heating and opposite-side heating.
Fig. \ref{fig:(i)-TB-elastic}(i) depicts the changes in effective
elastic coefficients as the thermal borders rise.

Conversely, Fig. \ref{fig:(i)-TB-elastic}(ii) specifically represents
the scenario of two-side heating, where there are two thermal boundaries.
The effective elastic coefficient $C_{11eff.}$ demonstrates identical
values at a volume fraction of 1.77 \% for both opposite and adjacent
wall heating. This is due to the uniform generation and distribution
of thermal stress in a single piezoelectric inclusion. However, at
a volume fraction of 5.33\%, the adjacent wall heating shows higher
negative values. This is because the top inclusion experiences more
thermal stress compared to the bottom two inclusions. In contrast,
in opposite wall heating, the temperature distribution is symmetric
for all inclusions, resulting in lower negative values. At a volume
percentage of 8.88\%, the piezoelectric distribution demonstrates
similarity in thermal boundaries. Additionally, the value of $C_{11eff.}$
remains equal for both adjacent and opposite wall heating. Moreover,
when the volume percent reaches 12.44\%, the system loses its symmetry,
resulting in a deviation in the observed value. However, when the
volume fraction reaches 16\%, the symmetry is restored, leading to
the same results as before. The thermal boundary conditions in this
scenario exhibit alternating symmetry and asymmetry concerning piezo-inclusion
distribution in the case of two-sided heating (BC1). Hence, the effective
elastic coefficients are contingent upon the alignment of piezoelectric
inclusions relative to the orientation of thermal barriers. The piezoelectric
inclusions exhibit symmetry about the orientation of thermal boundary
conditions in the context of two-side heating, specifically in cases
of adjacent and opposite wall heating, concerning $C_{33eff.}$ and
$C_{13eff.}$. Hence, it is evident from Fig. \ref{fig:(i)-TB-elastic}(ii(b\&c))
that the coefficients demonstrate comparable values regardless of
variations in the orientation of piezo-inclusions and thermal boundary
conditions. This consistency arises due to the symmetrical orientation
of the piezo-inclusions in accordance with thermal boundary conditions
(BC2) in the given scenario.

Hence, it can be inferred that the thermal boundaries' application
and alignment have an impact on the thermo-elastic characteristics
of the material. Nevertheless, this influence is not enduring and
dissipates once the thermal boundary conditions are removed. The thermo-elastic
and piezoelectric behaviour of a material, as well as its output responses,
are influenced by the formation and degradation of thermal stress
resulting from both external and internal heating. These thermal stress-induced
changes have a significant impact on the effective elastic characteristics
of the material.

\subsubsection{Effects of temperature regulation on mechanical output parameters:}

\begin{figure}
\includegraphics[scale=0.15]{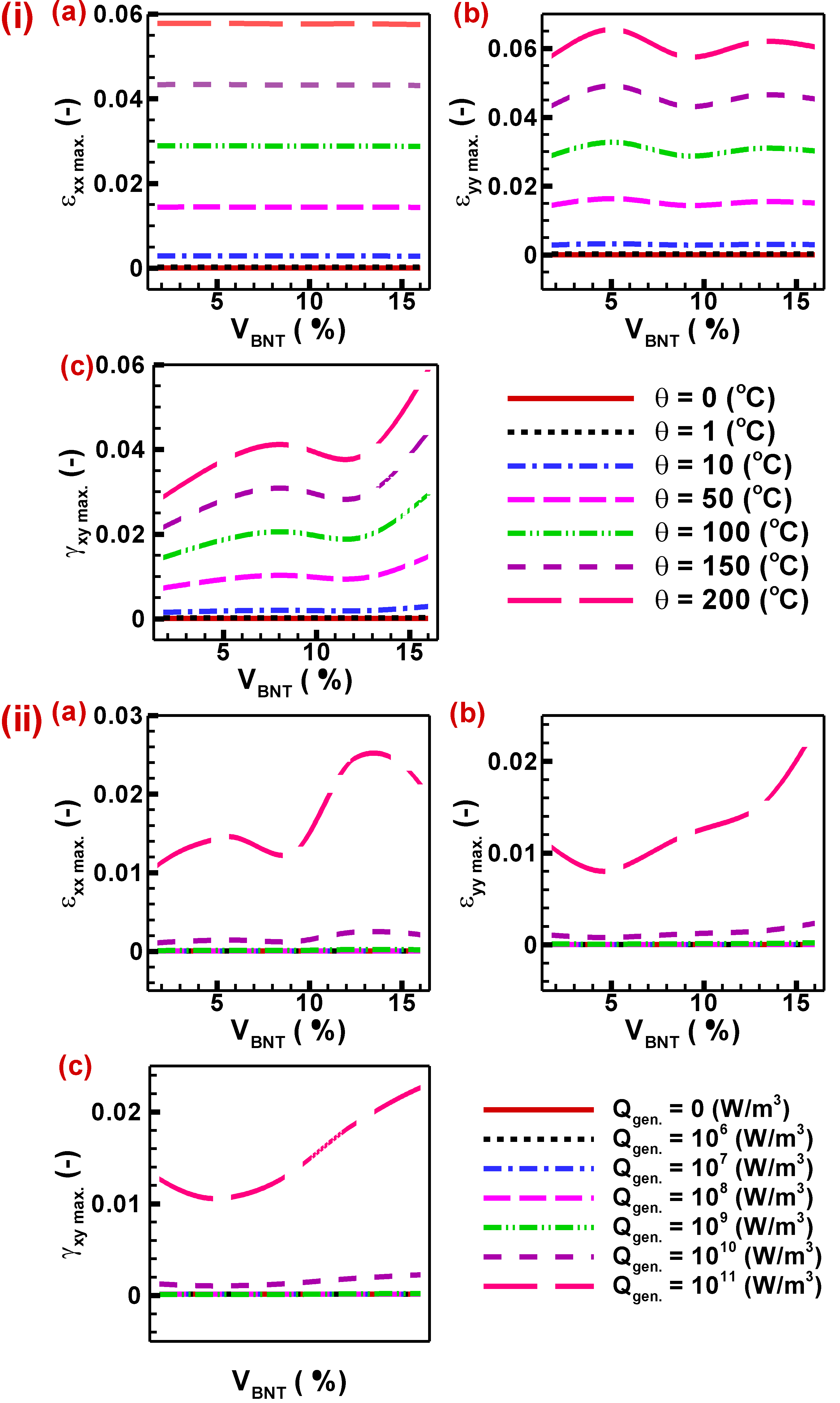}

\caption{(i) Effect of one side heating on (a) $\varepsilon_{xx_{max.}}$,
(b) $\varepsilon_{yy_{max.}}$, and (c) $\gamma_{xy_{max.}}$. (ii)
Effect of internal heat generation on (a) $\varepsilon_{xx_{max.}}$,
(b) $\varepsilon_{yy_{max.}}$, and (c) $\gamma_{xy_{max.}}$.\label{fig:(i)Mechanical-parameters}}

\end{figure}

The mechanical output properties of the BNT-PDMS matrix are influenced
by the application of thermal boundary conditions. This is demonstrated
by the observed fluctuation in effective elastic coefficients with
changes in temperature. These characteristics include principal and
transverse strains. Fig. \ref{fig:(i)Mechanical-parameters} (i) illustrates
the temperature-dependent variations in (a) $\varepsilon_{xx_{max.}}$,
(b) $\varepsilon_{yy_{max.}}$, and (c) $\gamma_{xy_{max.}}$, under
the One-side heating condition. Here,$\varepsilon_{xx_{max.}}$ represents
the maximum principal strain along the x direction, $\varepsilon_{yy_{max.}}$
represents the maximum principal strain along the y direction, and
$\gamma_{xy_{max.}}$ represents the maximum transverse strain. The
principal strain exhibits a positive correlation with the elevation
of wall temperature, primarily attributable to the emergence of thermal
strain. Conversely, the thermal strain does not directly influence
the transverse strain; nevertheless, there exists a fluctuation in
mechanical stress resulting from the generation of thermal stress,
which consequently induces variability in the transverse strain.

Moreover, the effect of one-side heating can be visualized by examining
the maximum principal and transverse mechanical strains developed
within the matrix. The maximum principal strains exhibit similar behaviour,
but their magnitude differs due to different elastic properties in
different principal directions. There is negligible change in maximum
principal mechanical strains with volume fraction of piezoelectric
inclusions when one side heating is absent or present at a minimal
level (the magnitude of temperature difference is very small). However,
it is affected by the volume fraction of BNT-type piezoelectric inclusion
and its distribution as thermal strains develop and dominate near
the boundary where heating is applied, as shown in Fig. \ref{fig:(i)Mechanical-parameters}(i(a\&b)).
Therefore, the maximum principal strains start to exhibit a dip at
volume fraction of 5.33\%, which reaches a minimum value at 8.88\%,
after which it starts to increase and becomes constant at 12.44\%
and further due to uniform or symmetric distributions of BNT-type
piezo-inclusions. The amplitude of this dip rises with the temperature
due to the generation of more thermal strains near the hot boundary
and thermal stress where the edge is fully constrained. As mentioned
above, the elastic properties are different in different principal
directions. Therefore, this dip is tiny when the maximum principal
strain is measured in the x direction and has a significant value
in the y direction. The effect of non-uniform one-side heating on
transverse strain is visualized in Fig. \ref{fig:(i)Mechanical-parameters}(i(c)).
The variation of maximum transverse strain with volume fraction of
BNT-type piezo-inclusions is negligible in the absence of thermal
boundary conditions as there is variation only due to the mechanical
load applied and distribution of the inclusions in the proximity of
the strained boundary. The transverse strain is affected by the application
of thermal coupling. However, there is no impact of thermal stress
on shear stress. This variation in transverse strain is due to the
free expansion nature of any material and different constraints at
different geometry boundaries. This will affect the principal stresses
of the body, and the transverse strain will vary accordingly as a
result. The transverse strain is also dependent on the distribution
of piezo-inclusions. The data demonstrates a linear growth in value
until reaching a volume percentage of 8.88\%. Subsequently, there
is a decline in value until a volume fraction of 12.44\% is reached.
Following this, there is another increase in value until a volume
fraction of 16\% is attained. The peak's magnitude has a positive
correlation with temperature, indicating that higher temperatures
result in a greater influence of thermal stress on primary stresses.
The presence of these elevated amplitudes may also result in the mechanical
breakdown of materials at elevated temperatures. Nevertheless, the
thermoelectric effect is intensified in BNT-type piezoelectric materials
as a result of elevated temperatures, leading to an increased electric
field. The non-uniform temperature distribution inside the system
results in a non-uniform thermoelectric effect across the geometry.
This issue may be resolved by implementing a uniform heating mechanism
for the matrix.

Furthermore, the impact of internal heat generation on the magnitudes
of maximum principal and transverse stresses is illustrated in Fig.
\ref{fig:(i)Mechanical-parameters}(ii). The maximum principal and
transverse strains exhibit an upward trend as the amount of internal
heat generation rises. This phenomenon can be attributed to the heightened
thermal strain and stress that occur in the proximity of the BNT-type
piezo-inclusions. Nevertheless, the connection between the mechanical
parameters and the volume fraction of piezo-inclusions varies due
to changes in the effective elastic property of the matrix resulting
from the addition of piezo-inclusions, as well as alterations in the
symmetry of their distribution. The maximum principal strain in the
x direction first exhibits an upward trend until reaching a volume
fraction of 5.33\%. Subsequently, it experiences a decrease until
a volume fraction of 8.88\%. It then resumes an upward trend until
reaching a volume fraction of 12.44\%. However, beyond this point,
it once again undergoes a decrease until a volume fraction of 16\%.
The observed phenomenon may be attributed to the presence of both
symmetric and asymmetric distributions of piezo-inclusions concerning
the mechanical boundary conditions (BC1), as the thermal boundary
conditions (namely, internal heat production) for each piezo-inclusion
remain uniform. The magnitude of the oscillations in the curve is
positively correlated with the augmentation of internal heat generation.
The magnitude of variation in heat generation is minimal until reaching
a value of $10^{10}$ $W/m^{3}$ since the corresponding temperature
difference is minor. However, at a heat generation value of $10^{11}$
$W/m^{3}$, the shift becomes significant, and this effect can be
further amplified with increased heat generation. The maximum principal
strain in the y direction first demonstrates a decline until a volume
percentage of 5.33\% is reached, following which it progressively
increases as the volume fraction increases. The observed phenomenon
may be attributed to the presence of both symmetric and asymmetric
distributions of piezo-inclusions in relation to the mechanical boundary
conditions (BC2). It should be noted that the thermal boundary conditions,
namely the internal heat generation, remain consistent for each BNT
piezo-inclusion. The greatest transverse strain has a similar pattern
to the maximum primary strain in the y direction, owing to the exact
underlying cause. The influence of internal heat generation is relatively
negligible at lower temperature differences but becomes significant
at higher values. This results in more pronounced strains, which can
enhance the piezoelectric response. However, it is essential to note
that these higher strains can also cause mechanical failure in the
system due to distortion of the grain boundaries.

\begin{figure}
\includegraphics[scale=0.15]{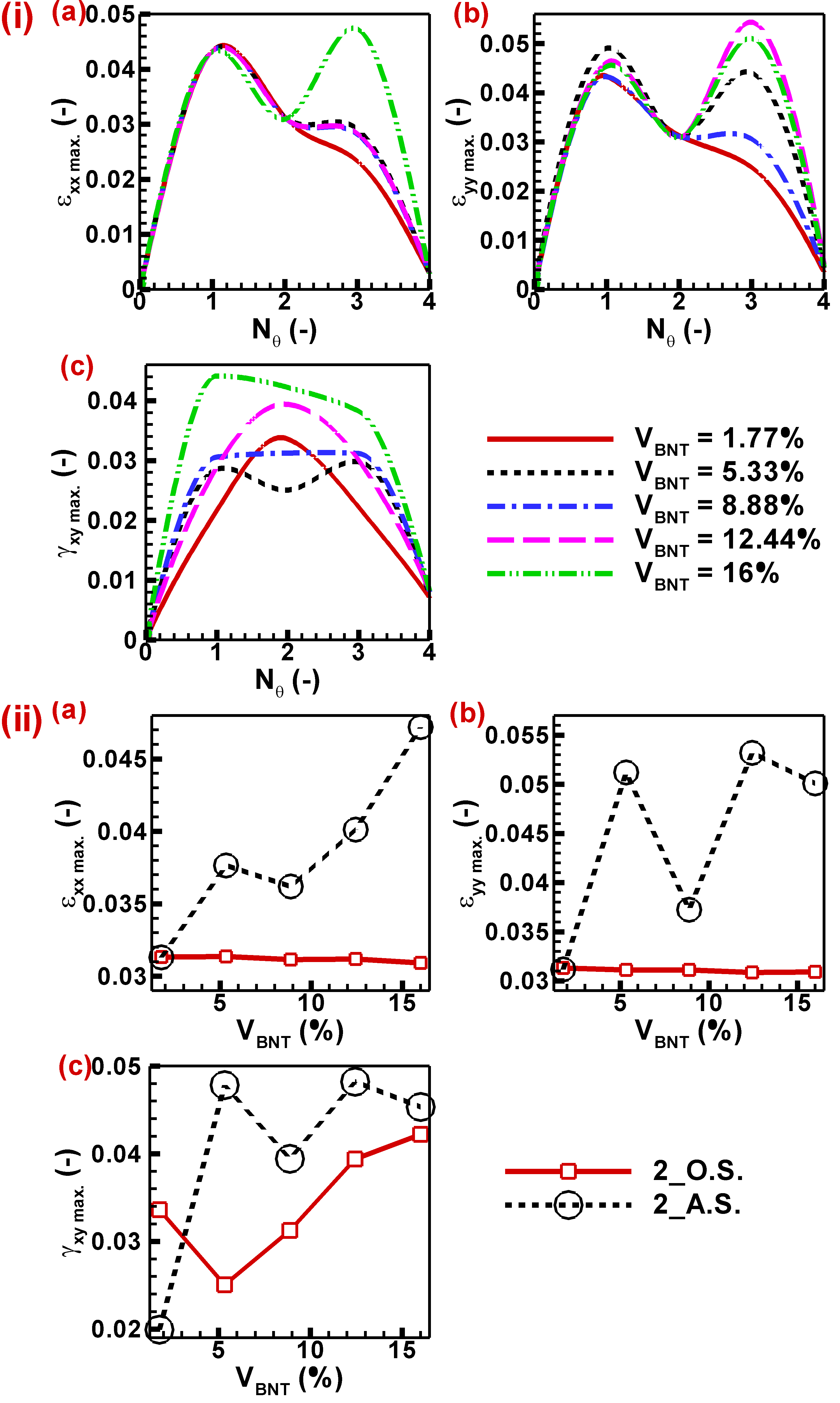}10\textasciicircum\{-3\}

\caption{(i) Effect of increase in several external thermal boundaries on (a)
$\varepsilon_{xx_{max.}}$, (b) $\varepsilon_{yy_{max.}}$, and (c)
$\gamma_{xy_{max.}}$. (ii) Comparison of (a) $\varepsilon_{xx_{max.}}$,
(b) $\varepsilon_{yy_{max.}}$, and (c) $\gamma_{xy_{max.}}$ for
two side adjacent wall heating and opposite wall heating.\label{fig:(i)TB-Mechanical_Parameters}}

\end{figure}

Moreover, the impact on mechanical parameters of the increase in thermal
boundaries in the case of external heating of the matrix has been
depicted in Fig. \ref{fig:(i)TB-Mechanical_Parameters} (i). The positive
maximum principal strain, which is on the order of $10^{-6}$, has
a negligible magnitude and is in close proximity to zero throughout
all volume fractions of BNT-type piezoelectric inclusions when there
are zero thermal boundaries present. This indicates only electromechanical
coupling. By introducing a single thermal boundary, specifically in
the scenario of one-side heating, the maximum value of $\varepsilon_{xx_{max.}}$
is observed to grow as a result of thermal strain. This holds true
for all instances involving different volume fractions of piezoelectric
inclusions. The aforementioned point represents the maximum value
of $\varepsilon_{xx_{max.}}$ for VBNT at four specific percentages:
1.77\%, 5.33\%, 8.88\%, and 12.44\%. Subsequently, the value of $\varepsilon_{xx_{max.}}$
diminishes as a result of a decrease in temperature differential or
thermal stratification. At a$V_{BNT}$ value of 16\%, the maximum
value of $\varepsilon_{xx_{max.}}$ is achieved for three thermal
barriers, as the distribution of piezoelectric inclusions normalizes
thermal stratification. In contrast, the augmentation of thermal barriers
to a total of four results in a heightened thermal stratification,
leading to a fall in the maximum value of $\varepsilon_{xx_{max.}}$
to lower magnitudes, approximately on the order of $10^{-3}$. The
greatest primary strain in the y direction, $\varepsilon_{yy_{max.}}$,
has a comparable behaviour to that of $\varepsilon_{xx_{max.}}$ for
$V_{BNT}$ values of 1.77\% and 5.33\%. Similarly, for $V_{BNT}$
values of 8.88\%, 12.44\%, and 16\%, $\varepsilon_{yy_{max.}}$ exhibits
a similar behaviour as $\varepsilon_{xx_{max.}}$ did for $V_{BNT}$$=16$\%.
The observed phenomenon can be attributed to thermal stratification
and piezo-inclusions' arrangement, considering specific mechanical
and thermal boundary conditions. The maximum transverse strain, denoted
as $\gamma_{xy_{max.}}$, has a linear correlation with the expansion
of thermal barriers, resulting in an increase in its magnitude.

When the values of $V_{BNT}$ are 1.77\% and 12.44\%, the maximum
value of $\gamma_{xy_{max.}}$ first increases until the number of
thermal barriers approaches 2, after which it subsequently declines.
The variable $\gamma_{xy_{max.}}$ has a peak value at $N_{\theta}=1$
and displays a parabolic trend as a result of its symmetrical geometric
characteristics. In contrast, when $V_{BNT}$ is equal to 8.88\% and
16\%, it peaks at $N_{\theta}=1$ and demonstrates a gradual decline
or a near constancy in $\gamma_{xy_{max.}}$ for $N_{\theta}$ values
ranging from 1 to 3. Subsequently, it experiences a quick reduction.
This phenomenon can also be attributed to the spatial arrangement
of piezoelectric inclusions in accordance with the mechanical boundary
conditions imposed, denoted as BC1 and BC2. The case of $V_{BNT}=5.33$\%
also demonstrates a comparable fluctuation, however, with a little
decrease in the value of maximum transverse strain at $V_{BNT}=8.88$\%.
This may be attributed to the asymmetric distribution of piezo-inclusions
in relation to the mechanical and thermal boundary conditions. There
are two distinct orientations of boundaries for two-sided heating,
namely neighbouring wall heating and opposite wall heating, as illustrated
in Fig. \ref{fig:(i)TB-Mechanical_Parameters}(ii). The opposing wall
heating demonstrates a consistent strain distribution inside the matrix,
resulting in a nearly constant variation with volume percent BNT for
maximal primary strains. The decrease in thermal strain is observed
as the volume fraction increases, albeit with a minimal slope. However,
it should be noted that the distribution of thermal strain is neither
uniform nor symmetrical concerning the piezo-inclusions. Consequently,
a zigzag pattern of maximum principal strains is obtained as the volume
fraction varies in accordance with the distribution of piezo-inclusions
within the matrix. The maximum transverse strain seen in neighbouring
wall heating has a similar pattern to that of the maximum primary
strain, which may be attributed to the same underlying reasons. However,
in the case of opposite wall heating, the maximum transverse strain
initially declines until a volume percentage of 5.33\% is reached,
after which it begins to grow with increasing volume fraction. The
reason for this similarity is due to the fact that the specific situation
mentioned ($V_{BNT}=5.33$\%) has very asymmetric piezo-inclusion
distributions. The presence of an asymmetric distribution results
in strain fluctuation due to disparities in the elastic and thermo-elastic
characteristics of the matrix and inclusions.

Furthermore, the augmentation of thermal boundaries amplifies the
strain within the system, thereby facilitating an increase in piezoelectric
response. However, it is crucial to consider that such significant
fluctuations in strain can also result in mechanical failures. This
consideration should be taken into account when designing piezoelectric
devices intended for high-temperature applications.

\subsection{Effects of thermal boundary conditions on mechanical field}

The influence of thermal boundary conditions on the electric field
may be observed by alterations in the patterns of electric potential
distribution. These changes arise from variations in the piezoelectric
effect, which is induced by modifications in mechanical strain resulting
from the formation of thermal strain. Furthermore, the thermoelectric
effect becomes relevant when thermal boundary conditions are applied,
resulting in the creation of temperature fluctuation ($\theta$) within
the composite material. The alteration in electric potential induces
changes in the distribution of electric field lines within the system,
thereby modifying the effective electric displacement. Therefore,
it is necessary to investigate the modification of the effective piezoelectric
coefficient in both directions. Hence, this part provides a comprehensive
explanation of the impact of thermal coupling on characteristics of
the electric field.

\subsubsection{Electric potential distribution:}

\begin{figure}
\includegraphics[scale=0.5]{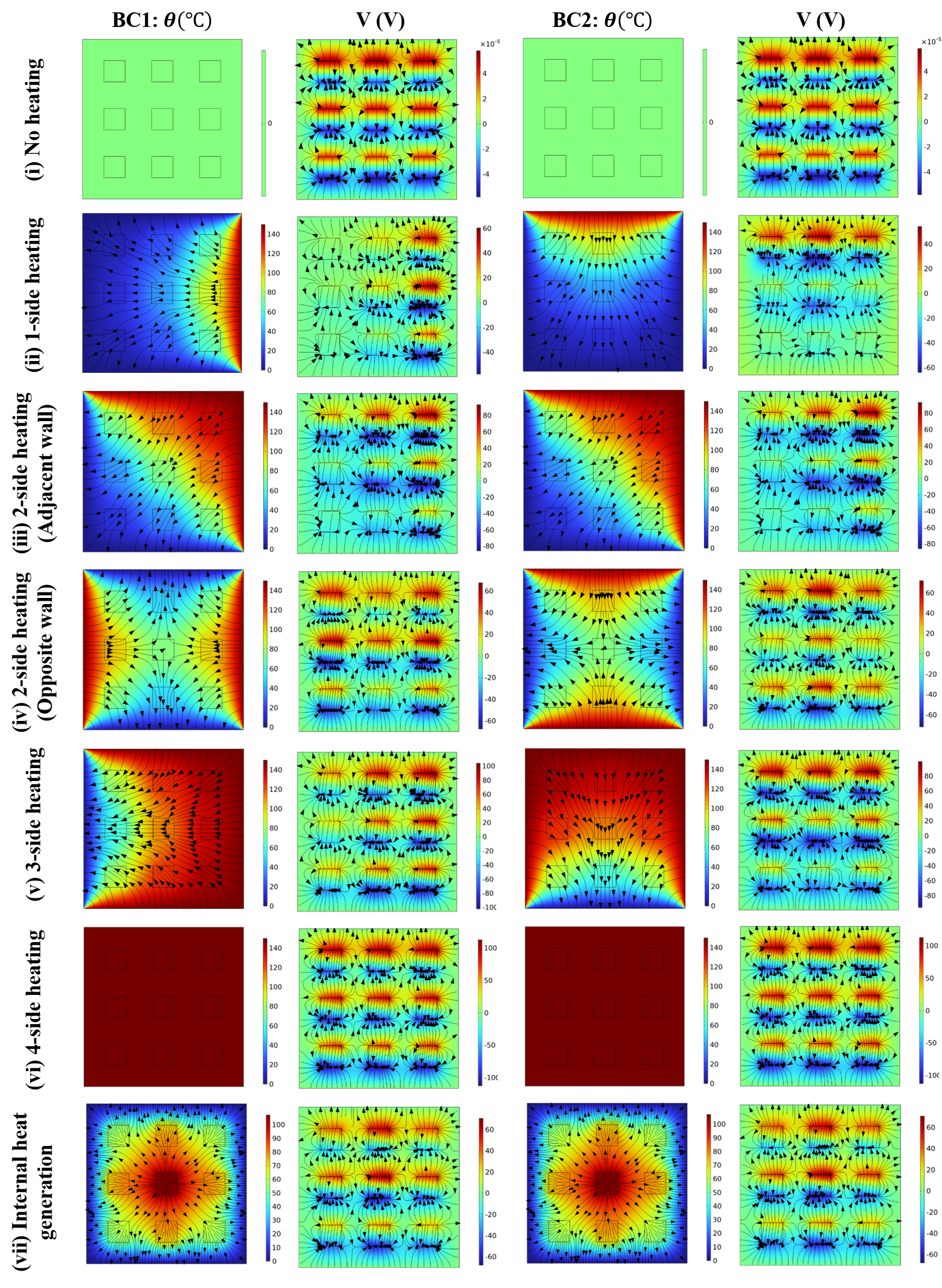}

\caption{Temperature and electric potential distribution contours for different
types of external heating (applied$\theta=150\lyxmathsym{\textcelsius}$)
and internal heat generation ($Q_{gen.}=10^{11}W/m^{3}$ ) for $V_{BNT}=16$
\% at mechanical boundary conditions BC1 \& BC2. Note that the vector
lines show the direction of heat transfer in the temperature contour
and electric field lines in the electric potential contour.\label{fig:Electric-potential}}

\end{figure}

The BNT-type piezoelectric-PDMS composite will behave differently
depending on the temperature and mechanical boundary conditions. The
heat field will modify the electrical field when thermal stresses
and strains occur, affecting the piezoelectric performance of the
materials. Changing the amplitude and direction of the thermal boundary
affects the heat transmission rate and temperature dispersion. Changes
in the number of thermal limitations will also affect the electrical
field. As a result, a study into the electrical behaviour of BNT-type
piezoelectric composites under different temperature boundary conditions
is critical. Fig. \ref{fig:Electric-potential} shows the contours
of temperature and electric potential distributions under various
boundary conditions. The heat transfer lines are drawn on the temperature
distribution curve to show the direction of heat transfer and heat
flow in the matrix and inclusions. When these lines enter inclusions,
their slope changes considerably, becoming steeper. As a result, the
heat transfer rate in the inclusions is faster than in the matrix,
and the temperature gradient in the inclusions is minimized, resulting
in a relatively uniform temperature within the inclusions. On the
other hand, the temperature gradient in the matrix and piezo-inclusions
in the direction of heat transport may be visualized. They will govern
the development of thermal stresses and strains in the geometry, affecting
the BNT material's piezoelectric response. The electric field lines
are overlaid on the electric potential distribution to determine the
intensity distribution of the electric field in the matrix.

Until there is no external or internal heating of the geometry, the
electric potential distribution, as illustrated in Fig. \ref{fig:Electric-potential}(i),
solely relies on mechanical stress. The electrical potential distribution
and electric field intensity are uniform along the BNT inclusions.
Thermal strain induction increased strain owing to thermal boundary
conditions, which will help strengthen the thermoelectric effect over
piezoelectric inclusions. The geometry is heated on one side in Fig.
\ref{fig:Electric-potential}(ii), and a thermal boundary condition
is imposed on the exact boundary where mechanical loading is applied.
In contrast, all other borders stay at room temperature. Since the
thermoelectric effect prevails near the hot wall and lessens towards
the cold wall, the total piezoelectric response improves as inclusions
near the hot wall display more robust thermoelectric behaviour. Increasing
the thermal boundary further enhances the electric distribution. The
mechanically laden fence and the neighbouring wall are heated, while
the other remains at room temperature (see Fig. \ref{fig:Electric-potential}(iii)).
As the high-temperature zone expands, so does the amount and dispersion
of electric potential. 

Due to the thermoelectric effect, maximum electric potential is found
at the boundary where heat is delivered, but only piezoelectric response
is recorded near the cool wall area. The strength of electric field
lines can be seen at the hot and cold wall regions, indicating that
intensity is higher near the hot wall due to improved thermoelectric
response and lower near the hard border due to lower thermoelectric
impact. In addition, the BNT-type piezoelectric composite is explored
for two opposing wall heating sides that exhibit a unique electric
potential distribution pattern, as shown in Fig. \ref{fig:Electric-potential}
(iv). Thermal boundary conditions are applied to both the mechanical
loading and constraint boundaries while the other walls remain at
ambient temperature. The electric distribution follows the temperature
distribution inside the geometry. The heat transfer direction is from
hot to cold walls, and heat transmission at the geometry's centre
is zero.

Similarly, the electric distribution is most extensive at both hot
walls. It diminishes as we move away from the walls due to electric
potential distribution, while inclusions at cool walls and the centre
of geometry display reduced thermoelectric impact. However, with this
type of thermal loading, the electric potential distribution for all
BNT piezo-inclusions is virtually uniform, as are the electric field
lines, which may benefit a consistent piezoelectric response.

Increasing the number of thermal boundaries to three expands the thermoelectric
effect distribution even more, as seen in Fig. \ref{fig:Electric-potential}(v).
In this situation, all three walls are heated except for the mechanical
constraint boundary, and heat is transmitted from these three walls
to the mechanical constraint border. The inclusions near the cold
walls have less thermoelectric effect, but the overall piezoelectric
distribution improves. Furthermore, in Fig. \ref{fig:Electric-potential}(vi),
the number of thermal boundaries is increased to four, and it is revealed
that no temperature gradient is observed inside the geometry, even
though the temperature of the entire geometry rises due to heating.
A consistent yet enhanced electric potential distribution was observed
over the whole geometry. The thermoelectric effect induced by all
side heating causes the increase in electric potential. The uniform
electric potential distribution indicates that the strength of the
electric field generated by temperature rise is constant in all directions.
The whole shape is intended to have a consistent and enhanced piezoelectric
response throughout all inclusions, albeit this may vary due to mechanical
loading.

We previously investigated the effect of non-uniform and uniform external
heating on the BNT-type piezoelectric composite. We found that an
increase in thermal barriers increases the spread of thermoelectric
effect distribution inside the geometry, which becomes uniform in
the case of constant heating. This is thought to improve piezoelectric
performance; however, the emphasis on internal heat production in
the electrical field has yet to be well investigated, as shown in
Fig. \ref{fig:Electric-potential}(vii). All boundaries are kept at
ambient temperature, and each piezoelectric inclusion generates heat
inside. The heat transmission from the source to the walls is uniform,
and the temperature distribution is the same, higher in the centre
and decreasing uniformly towards the wall; nevertheless, heat generation
from each inclusion enhances the temperature at that precise location.
Electric potential distribution is also the same, with a maximum in
the centre inclusion and a reduction as we travel towards the walls.
Although the electric potential is enhanced at each inclusion owing
to heating and the created thermal effect, the influence is less than
in the central inclusion. The most important characteristic is that
all inclusions have a more significant thermoelectric impact, which
is predicted to improve piezoelectric performance.

External or internal heating may positively influence the electric
field and increase the electric potential and intensity, resulting
in a better piezoelectric response. On the other hand, uniformly heated
surfaces offer a uniformly enhanced electrical field, which is more
beneficial for improving homogeneous piezoelectric response.

\subsubsection{Effects of thermal boundary conditions on effective piezoelectric
coefficients of materials:}

\begin{figure}
\includegraphics[scale=0.15]{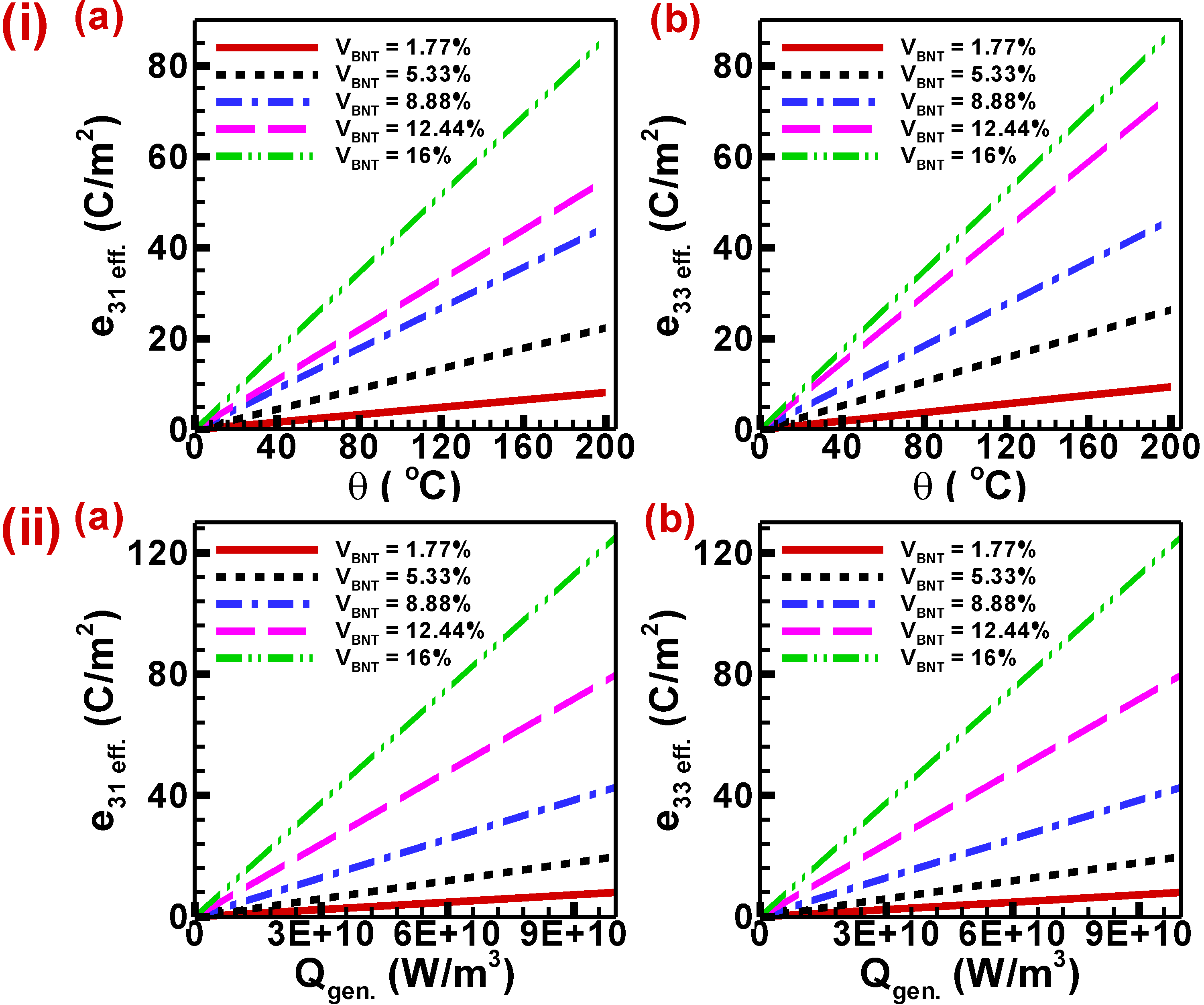}

\caption{(i) Effect of one side heating on effective piezoelectric properties
such as (a) $e_{31eff.}$, and (b) $e_{33eff.}$. (ii) Effect of internal
heat generation on effective piezoelectric properties such as (a)
$e_{31eff.}$, and (b) $e_{33eff.}$.\label{fig:Piezo_coeff}}

\end{figure}

The piezoelectric coefficients exhibit a linear relationship with
temperature difference, as seen in Fig. \ref{fig:Piezo_coeff}(i)
for the case of one-sided heating. Increasing the volume percentage
of BNT inclusions leads to a further enhancement in the piezoelectric
response. Given the aforementioned circumstances, thermo-electromechanical
coupling has an impact on the performance of piezoelectric materials.
The performance of piezoelectric materials is influenced by temperature
due to the presence of thermoelectric coupling, which becomes more
pronounced as the temperature rises, hence enhancing the piezoelectric
performance. The piezoelectric coefficient is enhanced when the temperature
increases due to the heightened thermal strain. This leads to an increase
in the overall strain, so generating a more resilient piezoelectric
response inside the inclusions. The application of mechanical strain
in the x-direction induces a piezoelectric response, namely the generation
of $e_{31eff.}$, in the z-direction. Despite the absence of symmetry
in the geometry of piezoelectric inclusions, both the effective $e_{31eff.}$
and $e_{33eff.}$ exhibit similar rates of growth in magnitude. This
phenomenon can be attributed to the uniform dispersion of thermal
stress in all orientations. Due to the diverse piezoelectric properties
and thermoelectric potential distribution in different orientations,
the magnitude of $e_{33eff.}$ is marginally greater than that of
$e_{31eff.}$. Moreover, the piezoelectric coefficients exhibit an
increase in effectiveness as temperature rises, mostly attributed
to the assistance provided by thermoelectric coupling. This phenomenon
becomes more pronounced at elevated temperatures and when a larger
proportion of the material consists of piezoelectric inclusions. Nonetheless,
it is crucial to consider the influence of phase shift and ferroelectric
domain switching on the amplification of the piezoelectric response
in order to accurately assess the true impact and calculate the threshold
value for temperature augmentation.

Fig. \ref{fig:Piezo_coeff}(ii) illustrates the linear relationship
between the piezoelectric coefficients and internal heat generation.
An augmentation in the piezoelectric response can be achieved by increasing
the volume percentage of BNT inclusions. This observation implies
that the performance of piezoelectric materials is influenced by thermo-electromechanical
coupling. The thermoelectric effects, which contribute to the enhancement
of piezoelectric performance, have a more substantial influence as
the amount of internal heat generation grows. This leads to non-uniform
temperature distributions inside the geometry and subsequently affects
the performance of the piezoelectric material. The enhancement of
the piezoelectric coefficient in the inclusions may be attributed
to the escalation of thermal strain resulting from an increase in
internal heat generation. This heightened thermal strain leads to
an overall increase in strain, so generating a more resilient piezoelectric
response. The mechanical strain applied in the x-direction generates
the piezoelectric response (a) $e_{31eff.}$ in the z-direction. Even
though the geometry lacks symmetry of piezoelectric inclusions, the
effective $e_{31eff.}$, and $e_{33eff.}$. display identical magnitude
increase rates. This is owing to the similar distribution of temperature
leading to symmetric thermal stress in all directions. Because the
material only has varied piezoelectric capabilities in various directions
and the thermoelectric potential distribution is uniform, the magnitude
of $e_{33eff.}$ is somewhat larger than (a) $e_{31eff.}$. Furthermore,
the effective piezoelectric coefficients rise with internal heat generation
due to thermoelectric assistance through thermoelectric coupling,
which becomes more dominant at higher values of internal heat generation
and a more significant volume percentage of piezoelectric inclusions.
However, the amplification of piezoelectric response is also sensitive
to phase shift and ferroelectric domain switching, which must be addressed
to determine the actual impact and internal heat generation threshold
value.

\begin{figure}
\includegraphics[scale=0.15]{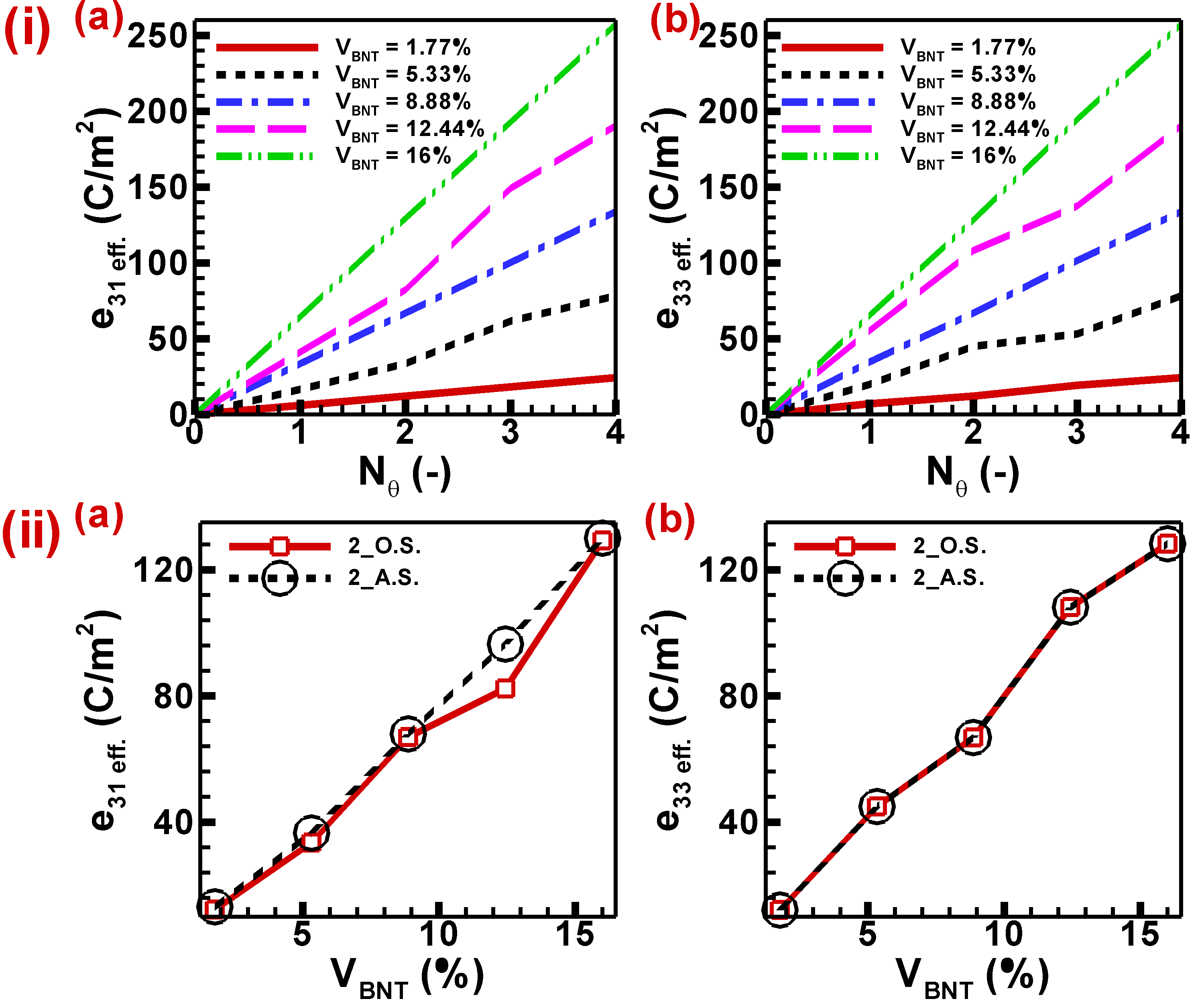}

\caption{(i) Effect of increase in number of external thermal boundaries on
effective piezoelectric properties such as (a) $e_{31eff.}$, and
(b) $e_{33eff.}$. (ii) Comparison of effective piezoelectric properties
such as (a) $e_{31eff.}$, and (b) $e_{33eff.}$ for two side adjacent
wall heating and opposite wall heating.\label{fig:TB-Piezo}}

\end{figure}

Moreover, Fig. \ref{fig:TB-Piezo}(i) illustrates the influence of
increased thermal boundaries on the effective piezoelectric coefficients,
which exhibit a positive correlation with the augmentation of thermal
boundaries. The effective elastic constants $e_{31eff.}$, and $e_{33eff.}$
display a linear relationship with the number of thermal barriers,
characterized by a consistent slope for volume fractions of 1.77\%,
8.88\%, and 16\%, correspondingly. The piezoelectric coefficients
exhibit a change in slope as the temperature limits increase, with
volume fractions of BNT at 5.33\% and 12.44\%. This shift is attributed
to the asymmetric distribution of piezo-inclusions. The disparity
between the effective piezoelectric coefficients $e_{31eff.}$, and
$e_{33eff.}$ is minimal due to the slight variations in the piezoelectric
characteristics of BNT in various directions. The piezoelectric coefficient
demonstrates enhanced effectiveness when the thermal boundaries are
expanded, resulting in an amplified thermoelectric effect caused by
an escalation in temperature disparity. Consequently, the thermoelectric
effect is amplified by the increasing volume percentage of BNT. Moreover,
there are two distinct orientations for two-sided heating, namely
neighbouring wall heating and opposite wall heating. The two distinct
orientations of dual-side heating have varying impacts on the effective
piezoelectric coefficients. The variation of the effective permittivity
$e_{31eff.}$ is influenced by the divergence in mechanical and thermal
boundary conditions relative to the poling direction of the piezoelectric
material. The value of $e_{31eff.}$ remains consistent for both adjacent
and opposite wall heating scenarios due to the symmetric nature of
the piezo-inclusion distribution under certain mechanical and thermal
boundary conditions. This holds true for $V_{BNT}$ values of 1.77\%,
8.88\%, and 16\%. The $V_{BNT}$ values, namely 5.33\% and 12.44\%,
exhibit modest variations across the two orientations of two-side
heating. This discrepancy can be attributed to the asymmetric distribution
of piezo-inclusions. The heating of the opposite wall exhibits significantly
lower values compared to the heating of neighbouring walls, which
can be attributed to a more equal distribution of thermal strain.
Furthermore, the value of the effective dielectric constant $e_{33eff.}$,
remains same for both adjacent and opposite wall heating scenarios
since the poling direction aligns with the mechanical and thermal
boundary conditions as shown in Fig. \ref{fig:TB-Piezo}(ii). The
thermoelectric effect seen across each piezo-inclusion remains consistent
under both temperature conditions of two-side heating, specifically
in relation to the effective piezoelectric coefficient. The thermo-mechanical
coupling phenomenon leads to an enhancement of the effective piezoelectric
coefficients. Nevertheless, it is crucial to consider the potential
occurrence of mechanical failure resulting from grain boundary distraction
caused by excessive thermal stress.

\subsubsection{Effects of thermal boundary regulation on electrical output parameters:}

The variation of piezoelectric and elastic coefficients with temperature
indicates that the application of thermal boundary conditions influences
the electrical output qualities of the BNT-PDMS matrix. The effective
material characteristics are influenced by the thermal boundary condition,
and the addition of piezoelectric inclusions to the original matrix
further impacts these properties. Consequently, the influence of heat
coupling on electromechanical behaviour is not the only factor at
play; an elevation in the volume percentage of BNT inclusions also
has an effect. Furthermore, the performance evaluation of the BNT-PDMS
composite may be conducted by analyzing several output metrics, such
as electric field intensity and potential, in addition to considering
its effective material features.

\begin{figure}
\includegraphics[scale=0.2]{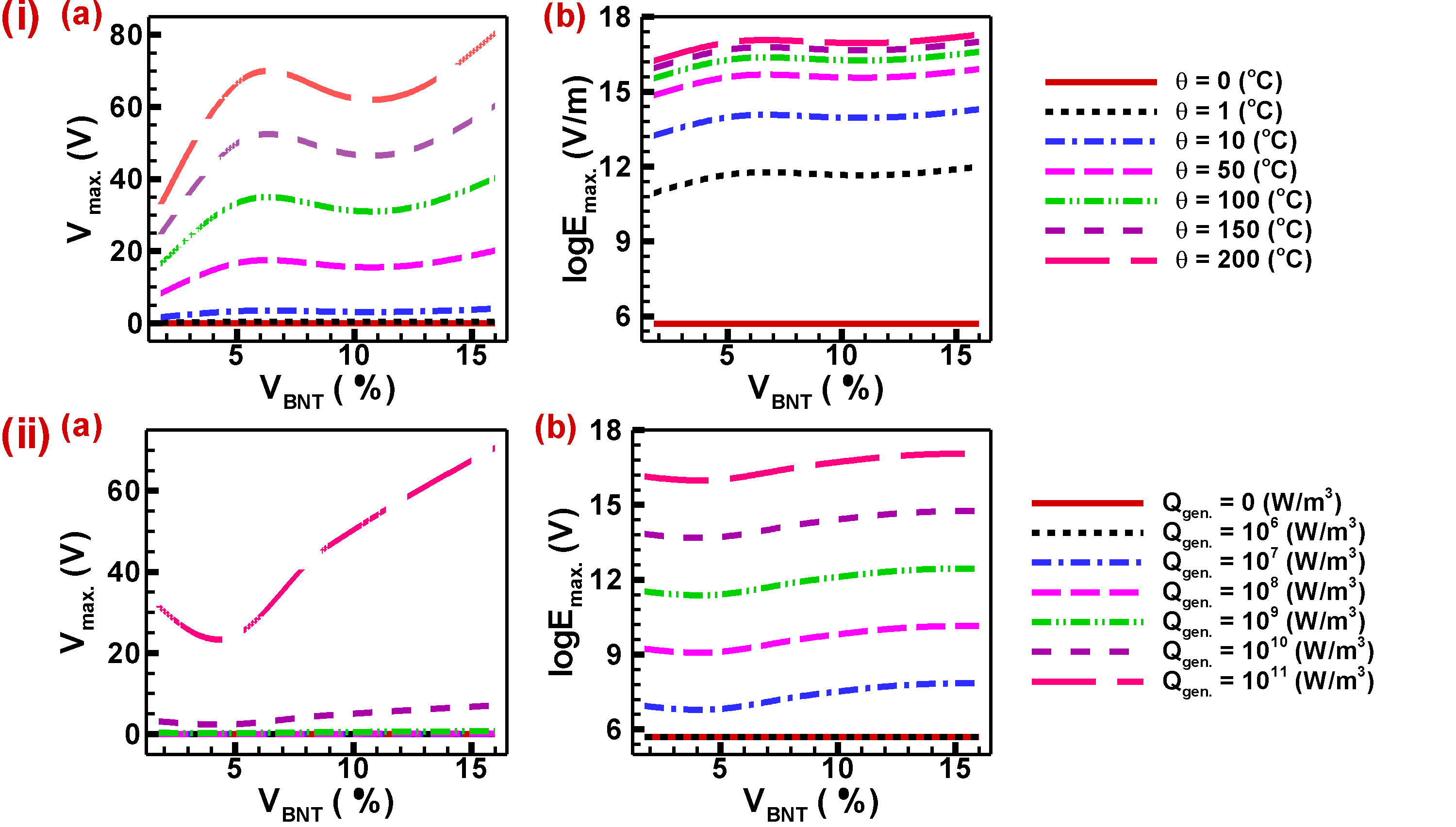}

\caption{(i) Effect of one side heating on (a) $V_{max}$, and (b) $\log E_{max.}$.
(ii) Effect of internal heat generation on (a) $V_{max}$, and (b)
$\log E_{max.}$.\label{fig:Electric-para}}

\end{figure}

Furthermore, the impact of unilateral heating on the electric output
characteristics may be seen by examining the variation of the maximum
electric potential with the volume \% of BNT-type piezoelectric inclusions
at various operating temperatures, as depicted in Fig. \ref{fig:Electric-para}(i(a)).
The increase in the volume percentage of piezo-inclusions leads to
an increase in the maximum potential generated in the piezoelectric
matrix without considering thermal coupling. However, the observed
increase is negligible since the electric potential is dispersed in
accordance with the doping material's geometry when piezoelectric
material is introduced. The electric potential is exclusively influenced
by the mechanical input that is applied. The application of a temperature
boundary condition on one side of the geometry results in the uniform
generation of the thermoelectric effect throughout all areas as the
temperature increases. Consequently, this leads to an increase in
the maximum electric potential. The amount of this increment has a
negative correlation with lower temperatures and a positive correlation
with higher temperatures. Nevertheless, the impact of volume fraction
on maximum electric potential is influenced by one-sided heating.
In contrast to the relatively moderate piezoelectric effects, the
presence of piezo-inclusions inside the region exhibits a more pronounced
thermoelectric influence. The variation of Vmax is also contingent
upon the distribution of piezo-inclusions. Initially, $V_{max.}$
shows an upward trend as the volume fraction grows, peaking at a volume
fraction of 8.88\%. Subsequently, $V_{max.}$ begins to decline, reaching
a global minimum at a volume fraction of 12.44\%. However, it subsequently
climbs once again, reaching a maximum value at a volume fraction of
16\%. The magnitude of $V_{max.}$ at a volume percent is significantly
greater than 1.77\%. Additionally, the thermo-electromechanical coupling
demonstrates substantial values of around 80V at a volume fraction
of 16\%, whereas it is on the order of $10^{-5}$ V when simply considering
electromechanical coupling. In the context of thermo-electromechanical
coupling, it is observed that the value of $V_{max.}$ exhibits numerous
rises in comparison to electromechanical coupling.

Figure \ref{fig:Electric-para}(i(b)) illustrates the relationship
between the maximum electric field intensity and the volume percent
of BNT inclusions at various operating temperatures. The logarithmic
scale is utilized to facilitate the representation and understanding
of the maximum electric field intensity. The incorporation of BNT
inclusions has resulted in an augmentation of the $\log E_{max.}$
value. Irrespective of the temperature boundary conditions, it can
be observed that the maximum electric field strength exhibits a linear
growth pattern in relation to the volume \% of piezo-inclusions. When
the temperature boundary condition is imposed on a single side of
the matrix, the magnitude of this fluctuation increases with the introduction
of one-sided heating. Within the region of elevated temperature, the
thermoelectric phenomenon surpasses the piezoelectric phenomenon,
resulting in an amplified enhancement amplitude when piezo-inclusions
are present. This overall increase in amplitude may be attributed
to the rise in temperature. The thermoelectric effect is influenced
by the distribution of inclusions inside the matrix due to the inconsistent
temperature distribution across the geometry. Consequently, the observed
phenomenon manifests itself in the form of variations in the intensity
of the electric field. The logarithmic value of the electric field
( $\log E_{max.}$) at $\theta=200\lyxmathsym{\textcelsius}$ is approximately
three times compared to the case when heating is absent, which indicates
that Emax. in thermo-electromechanical coupling (presence of different
thermal boundary conditions) is about 1000 times of $E_{max.}$ in
electromechanical coupling (absence of thermal boundary conditions)
of BNT-type piezoelectric composites.

Further, Fig. \ref{fig:Electric-para}(ii) exhibits the variation
of $V_{max.}$ and $\log E_{max.}$ with volume fraction at different
values of internal heat generation. The quantities first exhibit a
decline until reaching a volume fraction of 5.33\%. Subsequently,
they begin to climb and continue to do so as the volume percentage
further increases. Furthermore, it should be noted that the Vmax exhibits
a rise as the internal heat production grows. However, it is important
to acknowledge that this variation is rather little when considering
the modest temperature difference up to an internal heat generation
of $10^{10}$ $W/m^{3}$. Conversely, at an internal heat generation
of $10^{11}$ $W/m^{3}$, the value of $V_{max}$ becomes significantly
larger owing to the presence of larger theta values. Nevertheless,
the logarithmic representation of the maximum electric field intensity
($E_{max.}$) demonstrates a consistent increase in the presence of
internal heat generation. This uniform increment is due to the logarithmic
scale employed. If $E_{max.}$ were represented on an absolute scale,
the variance would not be uniform and would likely be far larger.
The voltage and electric field output of each piezoelectric inclusion
would exhibit a notable increase when subjected to increasing levels
of internal heat production, owing to the thermoelectric effect present
at each piezo-inclusion.

\begin{figure}
\includegraphics[scale=0.15]{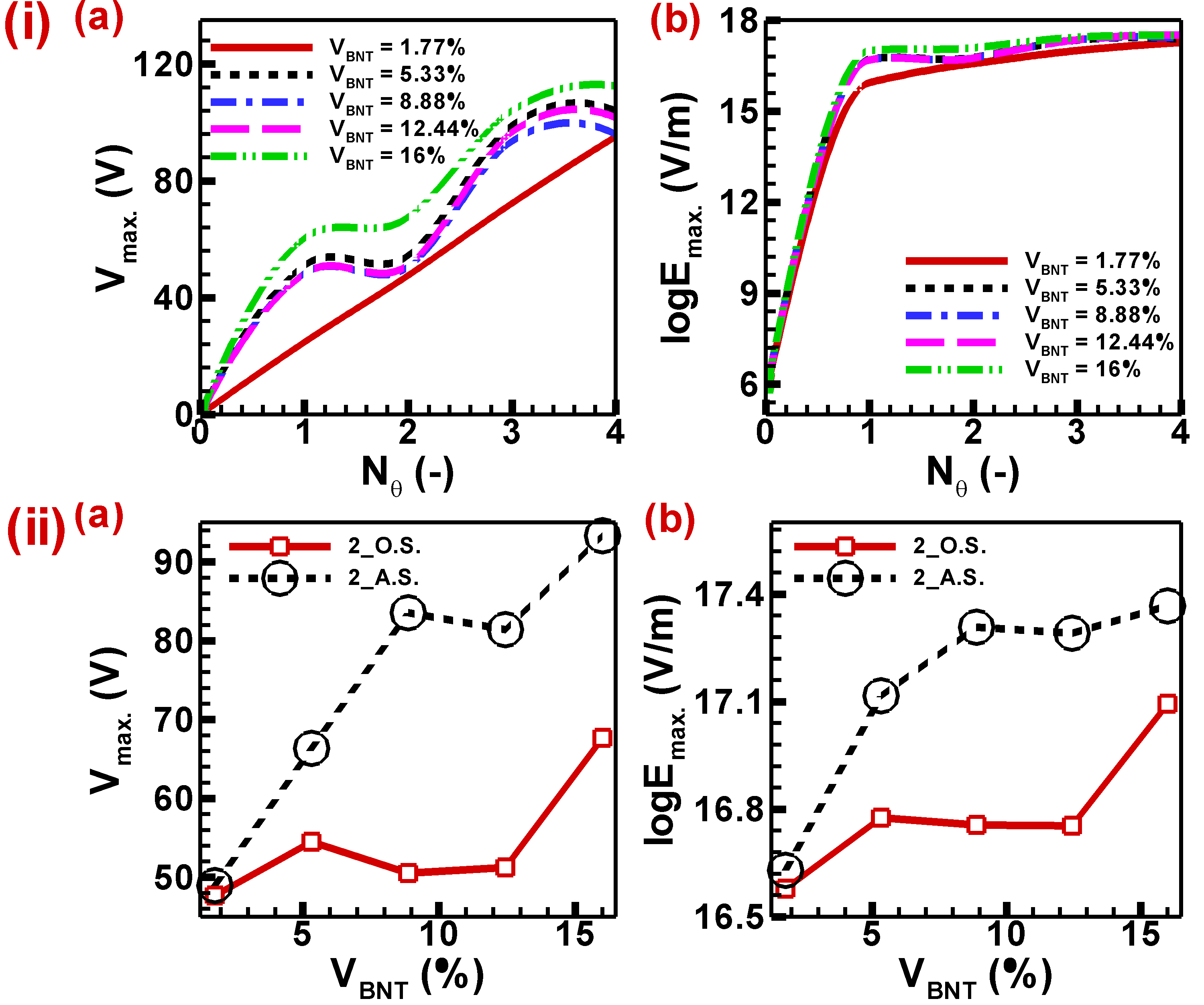}

\caption{(i) Effect of increase in several external thermal boundaries on (a)
$V_{max}$, and (b) $\log E_{max.}$ (ii) Comparison of (a) $V_{max}$,
and (b) $\log E_{max.}$ for two side adjacent wall heating and opposite
wall heating.\label{fig:TB-Electric-Para}}

\end{figure}

Moreover, Fig. \ref{fig:TB-Electric-Para}(i) exhibits the variation
of electric parameters such as $V_{max.}$ and $\log E_{max.}$ with
number of thermal boundaries. The maximum electric potential $V_{max.}$
first exhibits an upward trend as $N_{\theta}$ grows until $N_{\theta}$
reaches 1. Subsequently, Vmax remains constant from $N_{\theta}=1$
to 2 and then resumes its upward trajectory. In contrast, the logarithm
has a linear relationship with $N_{\theta}$ until $N_{\theta}=$1,
characterized by a steeper slope. Subsequently, the logarithm experiences
a growth with a diminished slope or rate. The observed phenomenon
can be attributed to the relationship between thermal boundaries and
thermal stratification. As the thermal boundaries increase, the thermal
stratification also increases. Consequently, the growth of electric
field intensity experiences a reduced slope. However, it is worth
noting that the electric field intensity exhibits a significantly
higher value when subjected to all-side heating boundary conditions
in comparison to other external and internal heating conditions. Moreover,
there exist two distinct orientations of two-sided heating boundary
conditions, namely neighbouring wall heating and opposite wall heating.
The heating of the neighbouring wall demonstrates a non-uniform distribution
of temperature differences within the geometry. This non-uniformity
leads to the formation of a thermoelectric effect, resulting in larger
values of $V_{max.}$ and $\log E_{max.}$ compared to the opposite
wall heating, as seen in Figure \ref{fig:TB-Electric-Para}(ii). The
observed variation in quantities related to adjacent and opposite
wall heating can be attributed to the asymmetric distribution of piezo-inclusions
along the direction of applied boundary conditions. This asymmetry
is also responsible for the zigzag pattern of increment in these quantities
with respect to the volume fraction at these two thermal boundary
conditions.

The phenomena of thermo-mechanical coupling results in an augmentation
of electric output characteristics for BNT-type composites. However,
it is essential to take into account the possible manifestation of
mechanical failure due to grain boundary distortion induced by extreme
thermal stress.

\subsection{Effects of flexoelectricity on mechanical and electric output parameters
accounting for thermo-electromechanical coupling:}

Flexoelectricity is a characteristic of dielectric materials with
an inherent electrical polarization that arises from a gradient in
strain. Flexoelectricity is a phenomenon that exhibits a close association
with piezoelectricity. However, it is essential to note that piezoelectricity
pertains to the polarization resulting from a uniform strain, whereas
flexoelectricity particularly pertains to the polarization arising
from strain variations across different points inside the material.
Flexoelectricity is a non-local size-dependent effect. In the context
of this investigation, the impact of flexoelectricity is considered
insignificant due to the relatively larger size of the piezo-inclusions
in comparison to the scale of flexoelectric coefficients associated
with the BNT materials.

\begin{figure}
\includegraphics[scale=0.15]{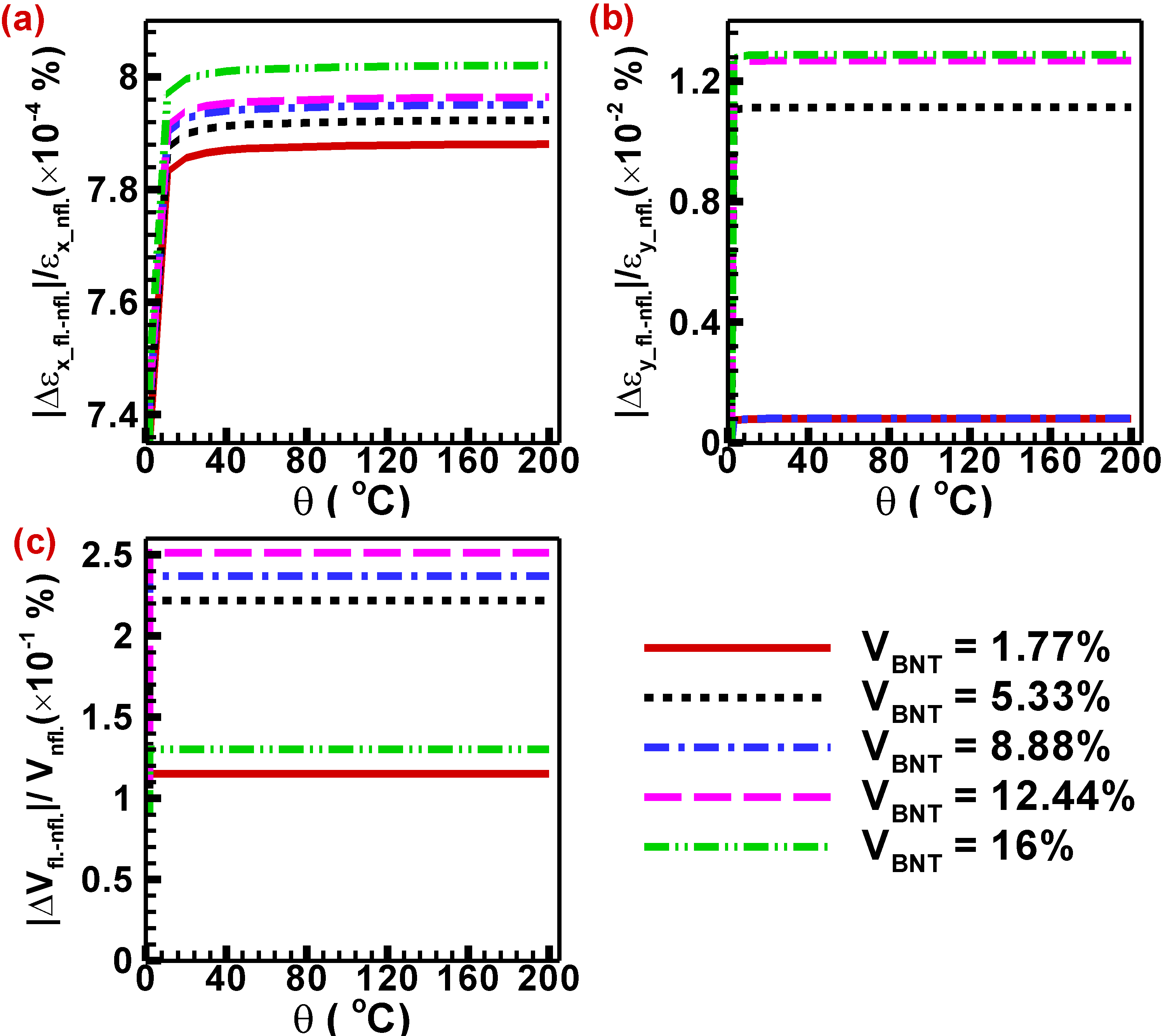}

\caption{Variation of mechanical and electrical output parameters such as (a)
\textgreek{e}\_x, (b) \textgreek{e}\_y, and (c) V for flexoelectric
and non-flexoelectric effects.\label{fig:Flexo}}

\end{figure}

Figure \ref{fig:Flexo} illustrates the relationship between temperature
and the mechanical and electrical output parameters, namely principal
strains and output voltage, for different volume fractions of BNT
inclusions. This depiction aims to examine the behaviour of BNT material
when subjected to flexoelectricity in the context of thermo-electromechanical
coupling. The research has revealed that the magnitude of major strain
in the x direction is around $10^{-4}$\%, while in the y direction
it is approximately $10^{-2}$\%. These values may be considered inconsequential.
Therefore, it can be concluded that the impact of flexoelectricity
on the mechanical domain is insignificant when considering piezo-inclusions
of this particular size. Moreover, the effect of flexoelectricity
on electric field parameters, such as electric potential, is around
$10^{-1}$\%, signifying a relatively little influence, yet greater
than that of the mechanical field. Based on the findings of this study,
it can be inferred that the impact of flexoelectricity is insignificant
for the given dimensions of the piezo-inclusions. Also, there is no
effect on increasing temperature difference has been observed at flexoelectricity
for this scale of piezo-inclusions. However, a potential increase
in the flexoelectric impact may be seen by reducing the size of the
piezo-inclusions. It is important to acknowledge that investigating
this aspect falls outside the scope of the present research.

\section{Conclusions}

A two-dimensional computational framework has been established to
examine the impact of thermo-electromechanical coupling on the performance
of lead-free BNT material when subjected to high-temperature haptic
applications. The thermo-electromechanical modelling of BNT ceramics
explicitly examines the macroscopic effects of phase changes on mechanical
and electric field parameters. The thermal stability of such material
must be evaluated due to its complex phase and domain structures,
which entirely depend on the material's temperature and grow even
more complicated at higher temperatures. The effective elastic and
piezoelectric properties and the mechanical and electric field parameters
of the two-dimensional framework have been investigated under various
temperature and mechanical boundary conditions. Based on the outcomes
of the study, several inferences can be drawn.
\begin{itemize}
\item The effective elastic coefficients exhibit higher negative values
for thermo-electromechanical coupling due to the generation of thermal
stress. 
\item The effect of the volume fraction of BNT inclusions on effective elastic
coefficients in the principal plane is compensated by thermal stress
for all thermal boundary conditions irrespective of mechanical boundary
conditions.
\item The considerable thermal strain is beneficial for a higher piezoelectric
response. However, there must be a limit as it can cause material
to fail mechanically due to grain boundary distortion. 
\item The electric field output parameters and effective piezoelectric coefficient
are enhanced due to the thermoelectric and piezoelectric effects. 
\item Among all thermal boundary conditions, uniform all-side heating (number
of thermal boundaries 4) exhibits the highest thermoelectric effect.
Electric field output values like electric potential and intensity
show enormous increments compared to electromechanical coupling. 
\end{itemize}
It is observed that compared to purely electromechanical case, BNT-type
piezoelectric composites behave differently when thermal field is
accounted consistently in the modelling main characteristics of these
lead-free materials. Therefore, the effect of thermo-electromechanical
coupling can\textquoteright t be neglected when designing lead-free
BNT- type piezoelectric materials for devices with thermal management
requirements, e.g., such as those used in haptic applications. The
effects of phase change and micro-domain dynamics have not been considered
in this study, where the main focus was to understand the effect of
temperature on the overall performance of BNT-based lead-free piezoelectric
composites based on the fully coupled thermo-electromechanical model.

\ack{}{}

The authors are grateful to the NSERC and the CRC Program (Canada)
for their support, and well as to the Ministerio de Ciencia e Innovacion
(Spain) through the research project PID2022-137903OB-I00. This research
was enabled in part by support provided by SHARCNET (www.sharcnet.ca)
and Digital Research Alliance of Canada (www.alliancecan.ca).

\bibliographystyle{unsrt}
\addcontentsline{toc}{section}{\refname}\bibliography{library}

\end{document}